\documentclass[fleqn,usenatbib]{mnras}

\usepackage{newtxtext,newtxmath}
\usepackage{adjustbox}
\usepackage{xcolor}

\usepackage[T1]{fontenc}

\DeclareRobustCommand{\VAN}[3]{#2}
\let\VANthebibliography\thebibliography
\def\thebibliography{\DeclareRobustCommand{\VAN}[3]{##3}\VANthebibliography}

\usepackage{graphicx}	% Including figure files
\usepackage{amsmath}	% Advanced maths commands
\usepackage{lineno}
\usepackage{eso-pic}% http://ctan.org/pkg/eso-pic

\AddToShipoutPictureBG*{%
  \AtPageUpperLeft{%
    \hspace{0.75\paperwidth}%
    \raisebox{-4.3\baselineskip}{%
      \makebox[0pt][l]{\textnormal{DES-2024-0848}}
}}}%

\AddToShipoutPictureBG*{%
  \AtPageUpperLeft{%
    \hspace{0.75\paperwidth}%
    \raisebox{-5.2\baselineskip}{%
      \makebox[0pt][l]{\textnormal{FERMILAB-PUB-24-0394-PPD}}
}}}%

\newcommand{\ddlr}{\ensuremath{d_\mathrm{DLR}}}
\newcommand{\msolar}{\ensuremath{\mathrm{M}_{\sun}}}
\title[SN Ia standardization and galactic location]{Reduction of the type Ia supernova host galaxy step in the outer regions of galaxies}

\author[Toy et al.]{
\parbox{\textwidth}{
\Large
M.~Toy,$^{1}$
P.~Wiseman,$^{1}$\thanks{E-mail:p.s.wiseman@soton.ac.uk}
M.~Sullivan,$^{1}$
D.~Scolnic,$^{2}$
M.~Vincenzi,$^{2,3,4}$
D.~Brout,$^{5}$
T.~M.~Davis,$^{6}$
C.~Frohmaier,$^{1}$
L.~Galbany,$^{7,8}$
C.~Lidman,$^{9,10}$
J.~Lee,$^{11}$
L.~Kelsey,$^{12,1}$
R.~Kessler,$^{13,14}$
A.~M\"oller,$^{15}$
B.~Popovic,$^{2}$
B.~O.~S\'anchez,$^{2,16}$
P.~Shah,$^{17}$
M.~Smith,$^{18}$
M.~Aguena,$^{19}$
S.~Allam,$^{20}$
O.~Alves,$^{21}$
D.~Bacon,$^{12}$
D.~Brooks,$^{17}$
D.~L.~Burke,$^{22,23}$
A.~Carnero~Rosell,$^{24,19,25}$
J.~Carretero,$^{26}$
L.~N.~da Costa,$^{19}$
M.~E.~S.~Pereira,$^{27}$
S.~Desai,$^{28}$
H.~T.~Diehl,$^{20}$
P.~Doel,$^{17}$
A.~Drlica-Wagner,$^{13,20,14}$
S.~Everett,$^{29}$
I.~Ferrero,$^{30}$
B.~Flaugher,$^{20}$
J.~Frieman,$^{13,20,14}$
J.~Garc\'ia-Bellido,$^{31}$
M.~Gatti,$^{11,14}$
E.~Gaztanaga,$^{7,12,8}$
G.~Giannini,$^{26,14}$
R.~A.~Gruendl,$^{32,33}$
G.~Gutierrez,$^{20}$
S.~R.~Hinton,$^{6}$
D.~L.~Hollowood,$^{34}$
K.~Honscheid,$^{35,36}$
D.~J.~James,$^{5}$
K.~Kuehn,$^{37,38}$
O.~Lahav,$^{17}$
S.~Lee,$^{39}$
J.~L.~Marshall,$^{40}$
J. Mena-Fern{\'a}ndez,$^{41}$
R.~Miquel,$^{42,26}$
A.~Palmese,$^{43}$
A.~Pieres,$^{19,44}$
A.~A.~Plazas~Malag\'on,$^{22,23}$
A.~K.~Romer,$^{45}$
S.~Samuroff,$^{46,26}$
E.~Sanchez,$^{47}$
D.~Sanchez Cid,$^{47,48}$
M.~Schubnell,$^{21}$
E.~Suchyta,$^{49}$
M.~E.~C.~Swanson,$^{32}$
G.~Tarle,$^{21}$
D.~L.~Tucker,$^{20}$
V.~Vikram,$^{50}$
A.~R.~Walker,$^{51}$
and N.~Weaverdyck$^{52,53}$
\begin{center} (DES Collaboration; Affiliations can be found at after the References.) \end{center}
}
\\
}

\date{Accepted XXX. Received YYY; in original form ZZZ}

\pubyear{\the\year{}}

\begin{document}
\label{firstpage}
\pagerange{\pageref{firstpage}--\pageref{lastpage}}
\maketitle

% Abstract of the paper
\begin{abstract}
Using 1533 type Ia supernovae (SNe Ia) from the five-year sample of the Dark Energy Survey (DES), we investigate the relationship between the projected galactocentric separation of the SNe and their host galaxies and their light curves and standardization. We show, for the first time, that the difference in SN Ia post-standardization brightnesses between high and low-mass hosts reduces from $0.078\pm0.011$\,mag in the full sample to $0.036 \pm 0.018$\,mag for SNe Ia located in the outer regions of their host galaxies, while increasing to $0.100 \pm 0.014$\,mag for SNe in the inner regions. The difference in the size of the mass step between inner and outer regions is $0.064\pm0.023$\,mag. In these inner regions, the step can be reduced (but not removed) using a model where the $R_V$ of dust along the line-of-sight to the SN changes as a function of galaxy properties. We investigate the remaining difference using the distributions of the SN Ia stretch parameter to test the inferred age of SN progenitors. Comparing red (older) environments only, outer regions have a higher proportion of high-stretch SNe and a more homogeneous stretch distribution. However, this effect cannot explain the reduction in significance of any Hubble residual step in outer regions. We conclude that the standardized distances of SNe Ia located in the outer regions of galaxies are less affected by their global host galaxy properties than those in the inner regions.

%To explain the remaining difference, we use the distributions of the SN Ia stretch parameter to test whether the inferred age of SN progenitors are more varied in the inner regions of galaxies. We find that the proportion of high-stretch SNe Ia in red (older) environments is more prominent in outer regions and that the outer regions stretch distributions are overall more homogeneous compared to inner regions, but conclude that this effect cannot explain the reduction in significance of any Hubble residual step in outer regions. We conclude that the standardized distances of SNe Ia located in the outer regions of galaxies are less affected by their global host galaxy properties than those in the inner regions. Old - too long (Marcus Change)
\end{abstract}


\section{Introduction}
\label{sec:introduction} 

Type Ia supernovae (SNe Ia) are a mature cosmological probe. The industrialisation of their discovery, measurement and analysis provides a \lq standardizable candle\rq\ giving unparalleled insight into the Universe's expansion history, particularly in the redshift $z<0.6$ Universe where dark energy dominates. However, despite the established consensus of an exploding carbon-oxygen white dwarf star with a light curve powered by the radioactive decay of $^{56}$Ni, there remain open questions over the details of the configuration of the progenitor systems, their explosion physics, and the effect of dust on their light curves and luminosities \citep[for reviews, see][]{Maguire2016,2023RAA....23h2001L}.

For cosmological applications, SN Ia brightnesses are standardized by the application of relations between SN Ia luminosity and SN light-curve width \citep{1993ApJ...413L.105P}, and SN Ia luminosity and SN optical colour \citep{1996ApJ...473...88R,1998A&A...331..815T}. It is well known that some SN Ia photometric properties used in this standardization have dependencies on the environment in which the SNe explode. For example, intrinsically brighter and slower evolving SNe Ia tend to explode in younger stellar populations \citep{2000AJ....120.1479H}. 

%For cosmological applications, such large scale environmental effects can be removed after the application of standardization relations between SN Ia luminosity and SN light-curve width \citep{1993ApJ...413L.105P}, and SN Ia luminosity and SN optical colour \citep{1996ApJ...473...88R,1998A&A...331..815T}.

More concerning for cosmological measurements is the observation that, even after such standardization, the brightnesses of SNe Ia (and thus the distances inferred to them) have a residual dependence on the properties of the galaxy in which they exploded. SNe Ia in massive, passive, older galaxies are brighter post standardization, and those in low-mass, younger, star-forming galaxies are fainter. Simplistically, this manifests as a step in SN Ia post-standardization luminosity at a host galaxy stellar mass close to 10$^{10}\,\msolar$ \citep{2010ApJ...715..743K,2010MNRAS.406..782S,2010ApJ...722..566L}. This so-called \lq mass step\rq\ is observed or routinely modelled in all large SN Ia surveys or compilations of datasets \citep[e.g.,][]{2014A&A...568A..22B,2022ApJ...938..110B,2023arXiv231112098R,2024arXiv240102929D,2024arXiv240520965G} and has a typical size of about 0.06--0.15\,mag (or three to seven per cent in distance) depending on the details of the sample. Similar trends are seen when replacing host stellar mass with other global host galaxy properties such as star-formation rate \citep[SFR;][]{2013A&A...560A..66R}, rest-frame colours \citep{2018A&A...615A..68R,2018ApJ...867..108J} or gas-phase/stellar metallicity \citep{2013ApJ...770..108C,2016ApJ...818L..19M,2022MNRAS.517.3312M}, and when considering these properties measured locally in a small aperture at the SN position \citep{2013A&A...560A..66R,2018A&A...615A..68R,2021MNRAS.501.4861K}.

The origin of the mass step is controversial and has implications that extend beyond SN Ia cosmology. Explanations include changing properties of the progenitor star, e.g., age \citep[][]{2019ApJ...874...32R,2020A&A...644A.176R} or metallicity \citep{2013ApJ...764..191H},  differing progenitor systems themselves, or dust properties that change with galaxy stellar mass \citep[][hereafter BS21]{2021ApJ...909...26B}. This latter explanation, where the ratio of total-to-selective extinction $R_V$ is different in low and high-mass galaxies, also provides an explanation for the variation of the step size with SN colour, with redder SNe appearing to show a larger step \citep{2021ApJ...909...26B}. At the time of writing, no single astrophysical origin for the mass step appears to explain all of the SN Ia observations \citep{2022MNRAS.517.2360T,2023MNRAS.520.6214W,2024arXiv240605051P}.

A complementary approach to investigating the mass step is to measure the effect that alternative standardisation procedures have on its size and significance. Examples of this include using near-infrared peak magnitudes \citep{2020ApJ...901..143U, 2021ApJ...923..197P, 2021ApJ...923..237J}, Bayesian hierarchical models \citep[e.g.][]{2022MNRAS.517.2360T}, or the spectral modelling of \lq twin\rq\ SNe Ia \citep{2021ApJ...912...71B}. Some of these approaches have led to a reduced mass step. \citet{2021ApJ...923..237J} found a mass step consistent with zero after allowing $R_V$ to vary between each SN in their light curve fit, although \citet{2022MNRAS.517.2360T} found that the step persisted regardless of their treatment of $R_V$. \citet{2021ApJ...912...71B} reduced the step from $0.092\pm0.024$ to $0.040\pm0.020$ when standardizing using a twins embedding approach. While promising in understanding the underlying cause of the step these methods all necessitate either expensive data and/or low-$z$ SNe Ia and cannot be used in the standardization of the current state-of-the-art large high-$z$ analyses.

In this paper, we explore whether the position of a SN Ia in its host galaxy can provide additional information on the mass step puzzle, in particular the projected separation of a SN
from its host galaxy (or galactocentric distance). There are strong astrophysical motivations for examining this in more detail: it has long been known that galaxies often show radial gradients in metallicity, age and dust content \citep[e.g.,][]{1971ApJ...168..327S,1981ARA&A..19...77P,1989ARA&A..27..235K,2019A&ARv..27....3M,2021MNRAS.502.5508P}. These may act as a proxy for the SN Ia progenitor star and its environment. For example, stellar populations in late-type galaxies are typically younger in the outer regions with a lower scatter in the ages measured  \citep{2017MNRAS.466.4731G,2021MNRAS.502.5508P}, whereas stellar populations in early-type galaxies may be older in the outer regions \citep{2007A&A...467..991B,2017MNRAS.466.4731G} or exhibit little overall gradient. Similarly, both star-forming \citep{1995ApJ...444..721S,2014A&A...563A..49S} and early-type galaxies \citep{1989ARA&A..27..235K} show negative gradients in their metal abundances \citep{2014ApJ...791L..16G,2015A&A...581A.103G}, with inner galactic regions being more metal rich than the outer regions.

Work using various samples of SNe Ia of order 100 events showed no clear trends between SN Ia light-curve width and their projected galactocentric distance \citep{2000ApJ...542..588I,2007ApJ...659..122J,2009ApJ...700.1097H,2010AJ....139...39Y}. Using a larger sample of $\sim1000$ events from the Dark Energy Survey five-year sample, \citet{2023MNRAS.526.5292T} showed a deficit of brighter and more slowly evolving (high \lq stretch\rq) SNe Ia in the centres of galaxies. This is unlikely to originate from an observational bias: the well-known increased difficulty of finding SNe Ia in galactic centres due to the increased surface brightness would not bias towards detecting the faster, fainter SNe that appear preferentially hosted in these regions. 

In terms of SN Ia colour (or extinction), there is some evidence that extincted (red) SNe Ia are not found at large galactocentric distances \citep{2007ApJ...659..122J,2009ApJ...700.1097H} and are instead found more centrally in spiral or star-forming galaxies \citep{2012ApJ...755..125G,2015MNRAS.448..732A}. Using 302 SNe Ia from the SDSS SN survey \citep{2018PASP..130f4002S}, \citet{2018MNRAS.481.2766H} showed a statistically significant difference in the observed SN Ia colour distributions between events that are nearer (redder SNe) and further (bluer SNe) from their host galaxy centres, confirmed in the recent release of $\sim$1000 low-redshift SNe Ia from the Zwicky Transient Facility \citep[ZTF;][]{2024arXiv240602072G}.

Studies into the cosmological effects of galactocentric distances on SN Ia distance determination have been limited. No clear results were found in the low-redshift sample of \citet{2009ApJ...700.1097H}. However, \citet{2018MNRAS.481.2766H} and \citet{2020ApJ...901..143U}  report a smaller residual scatter in SN Ia Hubble residuals at larger galactocentric distances. There is also some evidence that galactic position may influence other SN Ia properties. \lq High velocity\rq\ SNe Ia, defined as those SNe Ia with silicon velocities measured from their spectra in excess of 12,000\,km\,s$^{-1}$, are found preferentially in the centres of their host galaxies \citep{2013Sci...340..170W,2015MNRAS.446..354P,2023arXiv230410601N}.

We use the Dark Energy Survey (DES) SN programme five-year dataset \citep{2024arXiv240605046S} to examine the relationship between galactocentric distance and SN Ia standardization. The DES-SN5YR dataset is a well-calibrated and understood high-redshift SN Ia cosmological dataset, with deep imaging data available for the SN Ia host galaxies \citep{2020MNRAS.495.4040W}. Throughout, where relevant we assume a reference cosmology of a flat $\Lambda$CDM universe with $\Omega_\mathrm{matter}=0.3$, and with a Hubble constant of $H_0=70$\,km\,s$^{-1}$\,Mpc$^{-1}$.

\begin{figure*}
    \includegraphics[trim={1cm 1cm 0 0},clip, width=0.25 \textwidth]{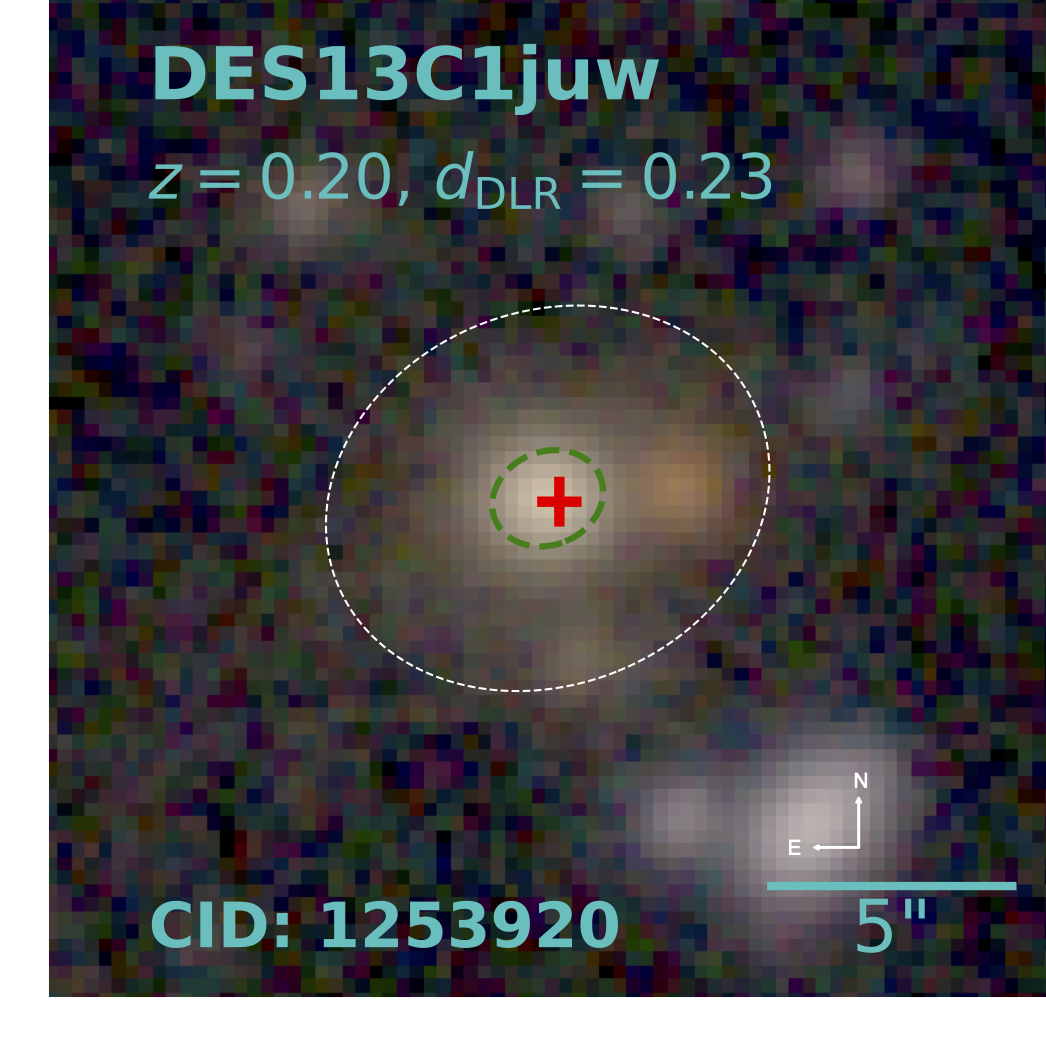}
    \includegraphics[trim={1cm 1cm 0 0},clip, width=0.25 \textwidth]{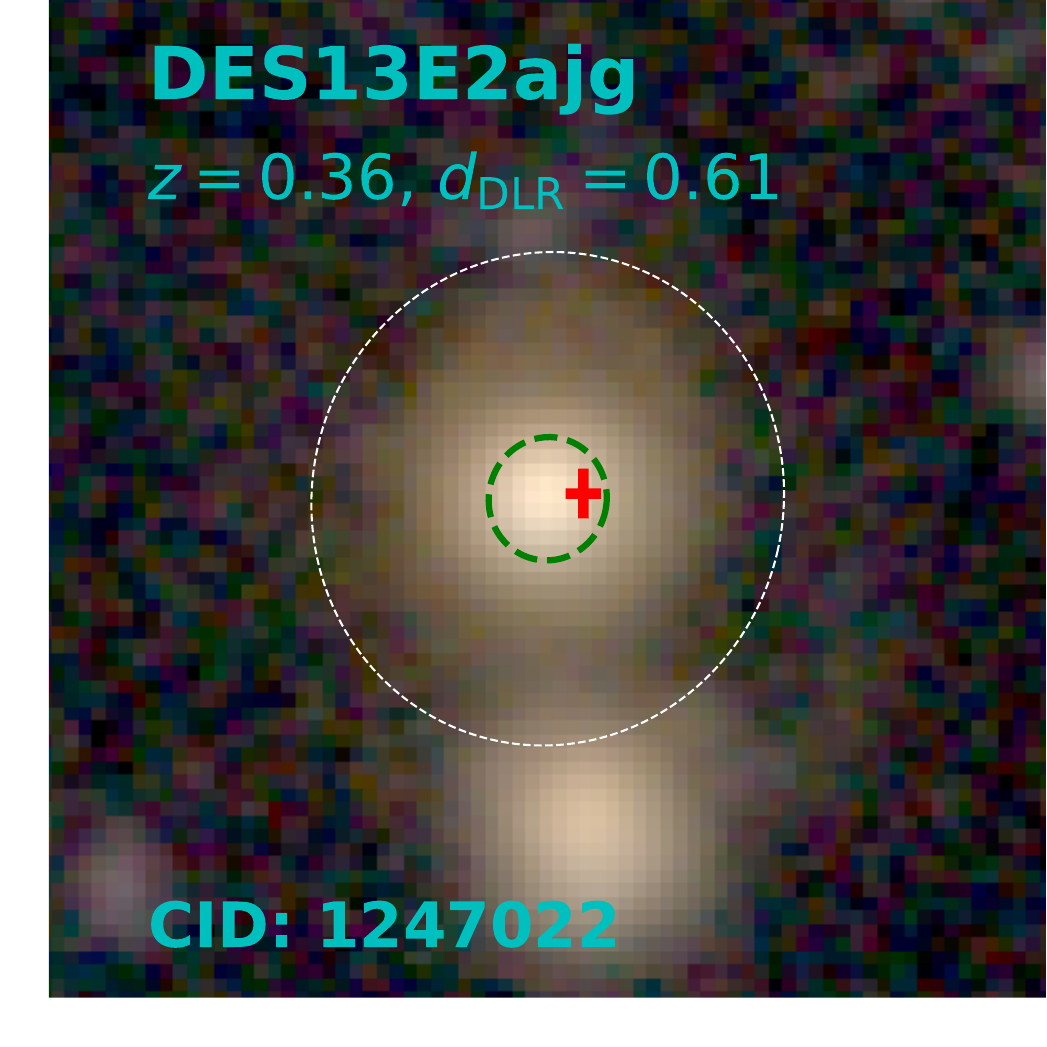}\includegraphics[trim={1cm 1cm 0 0},clip, width=0.25 \textwidth]{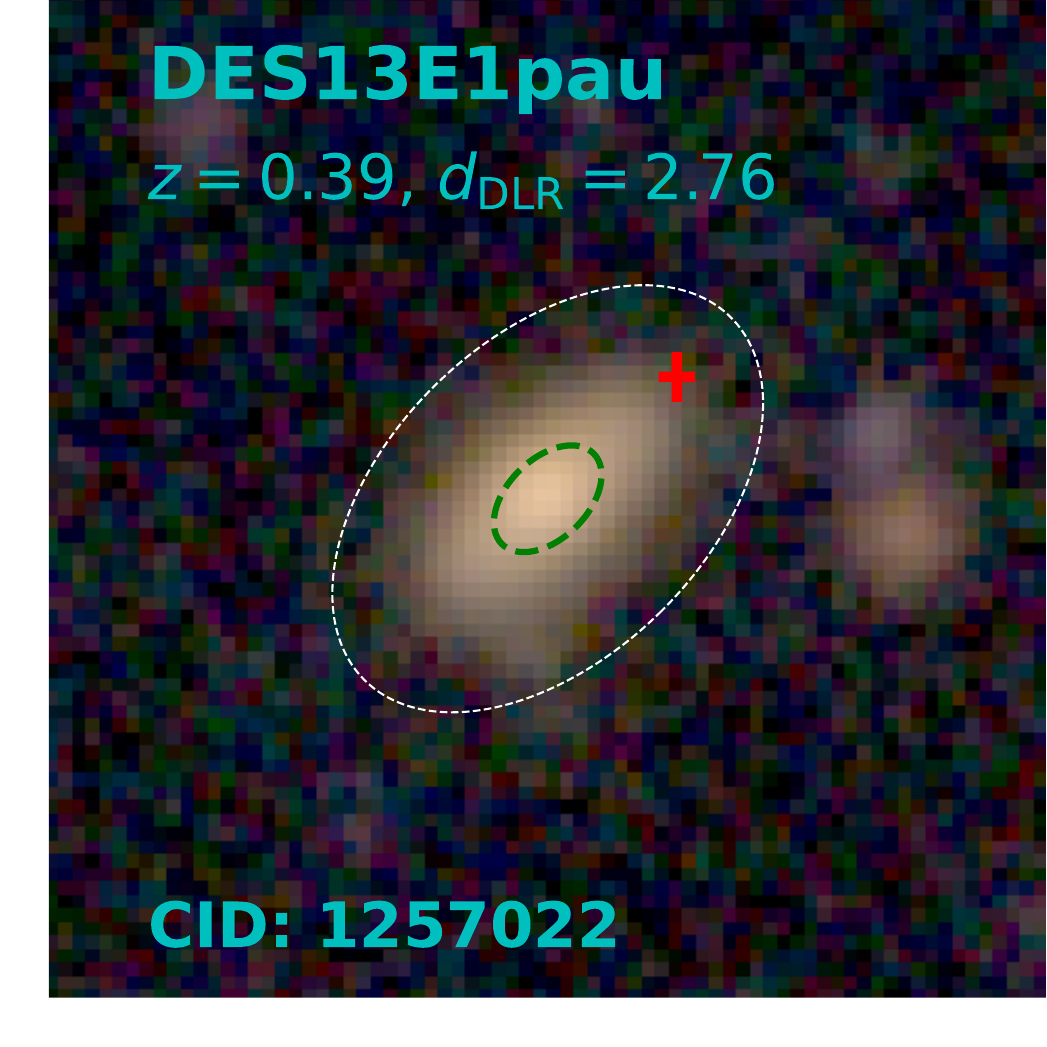}
    \includegraphics[trim={1cm 1cm 0 0},clip, width=0.25 \textwidth]{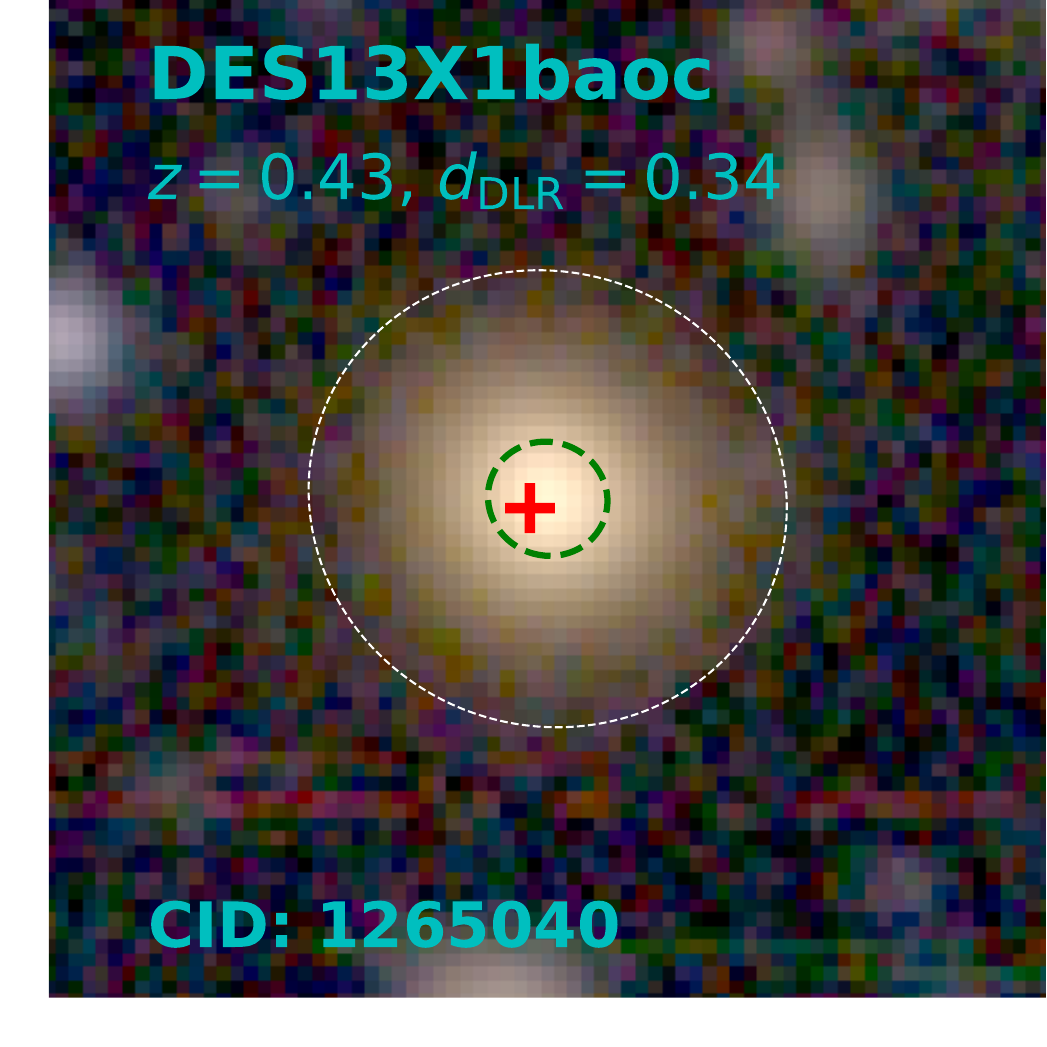}
    \includegraphics[trim={1cm 1cm 0 0},clip, width=0.25 \textwidth]{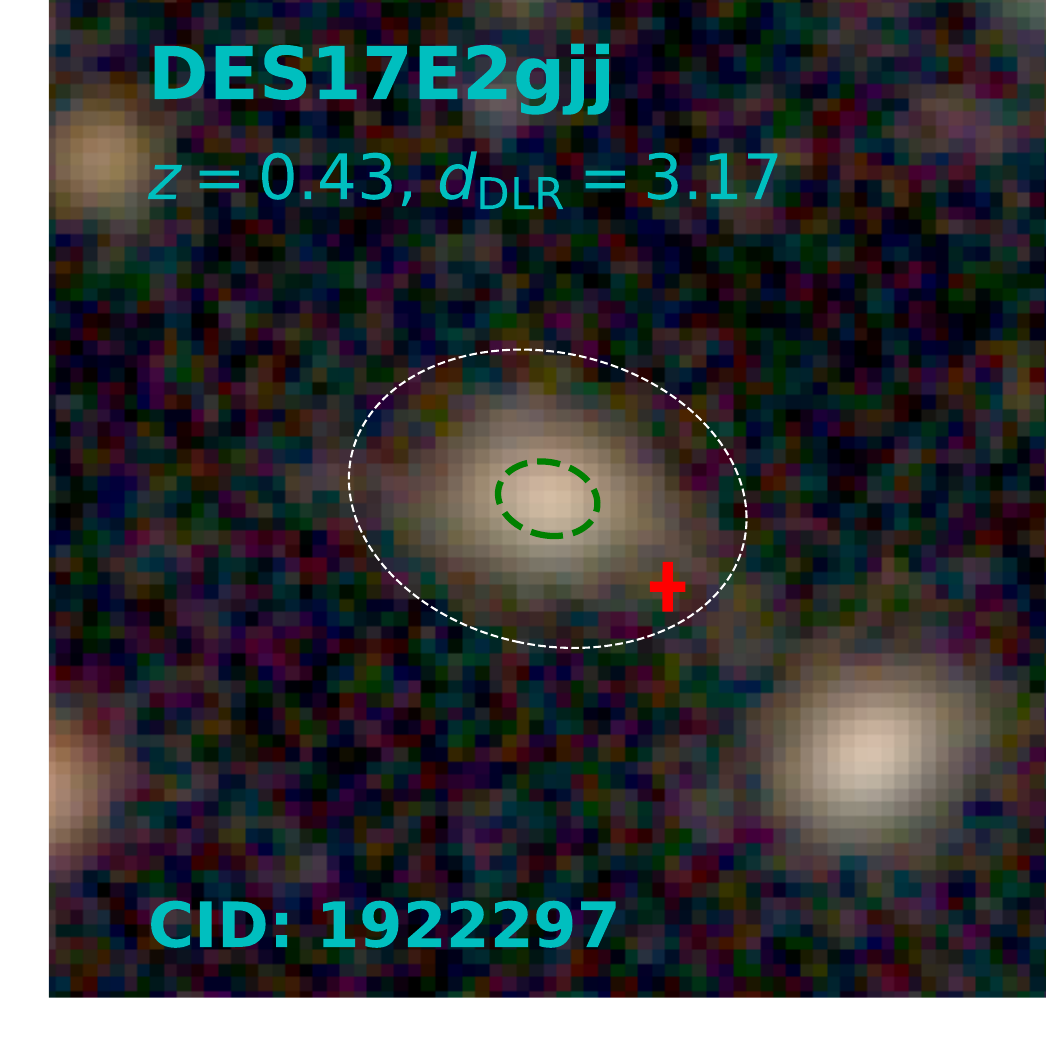}
    \includegraphics[trim={1cm 1cm 0 0},clip, width=0.25 \textwidth]{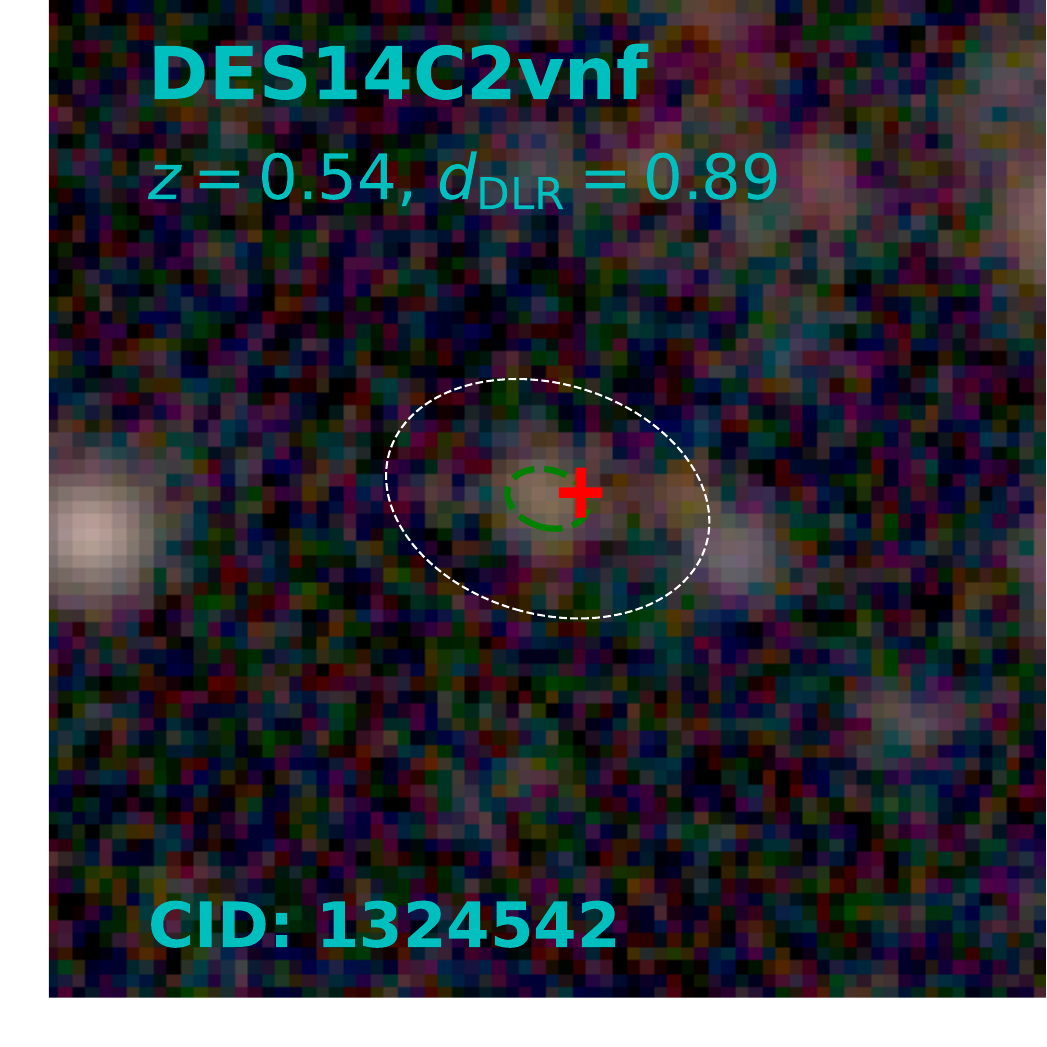}
    \includegraphics[trim={1cm 1cm 0 0},clip, width=0.25 \textwidth]{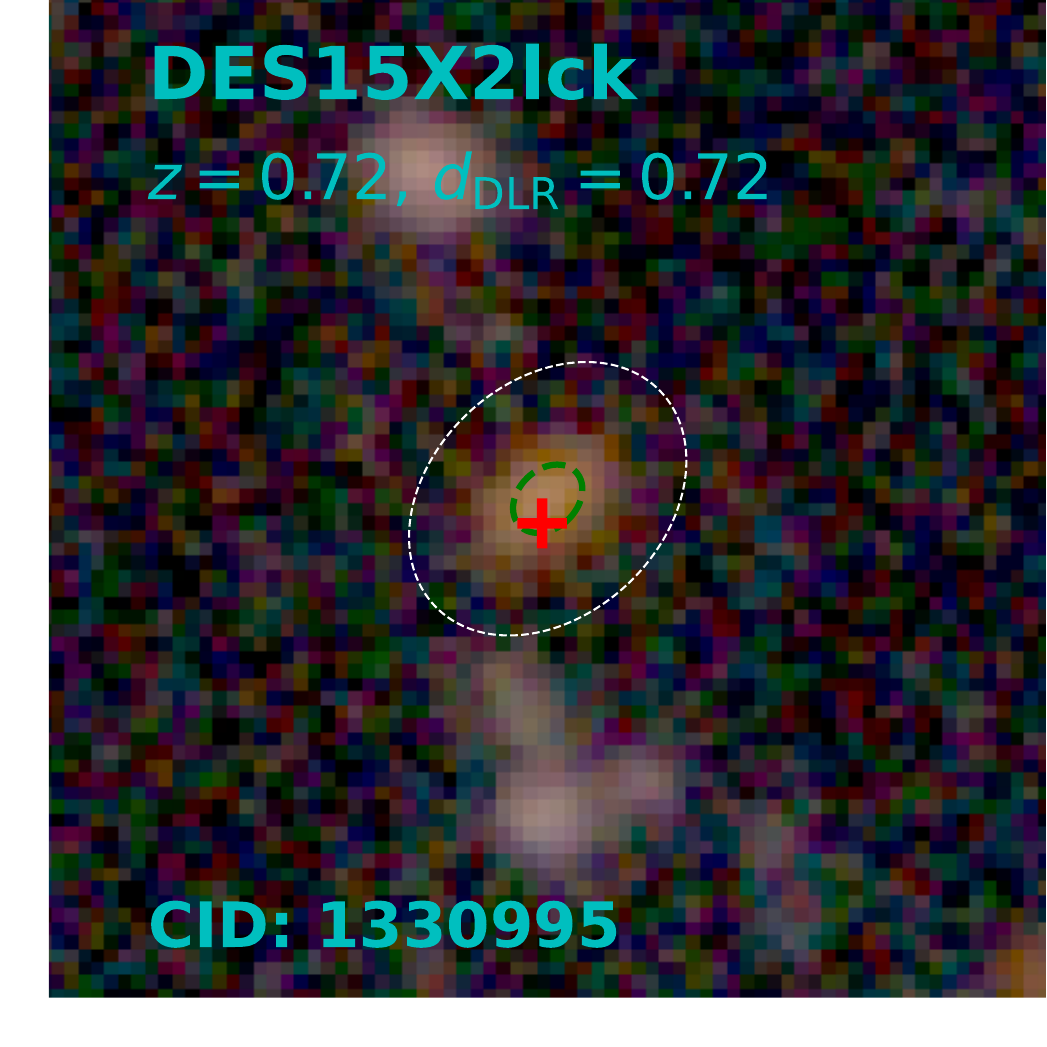}
    \includegraphics[trim={1cm 1cm 0 0},clip, width=0.25 \textwidth]{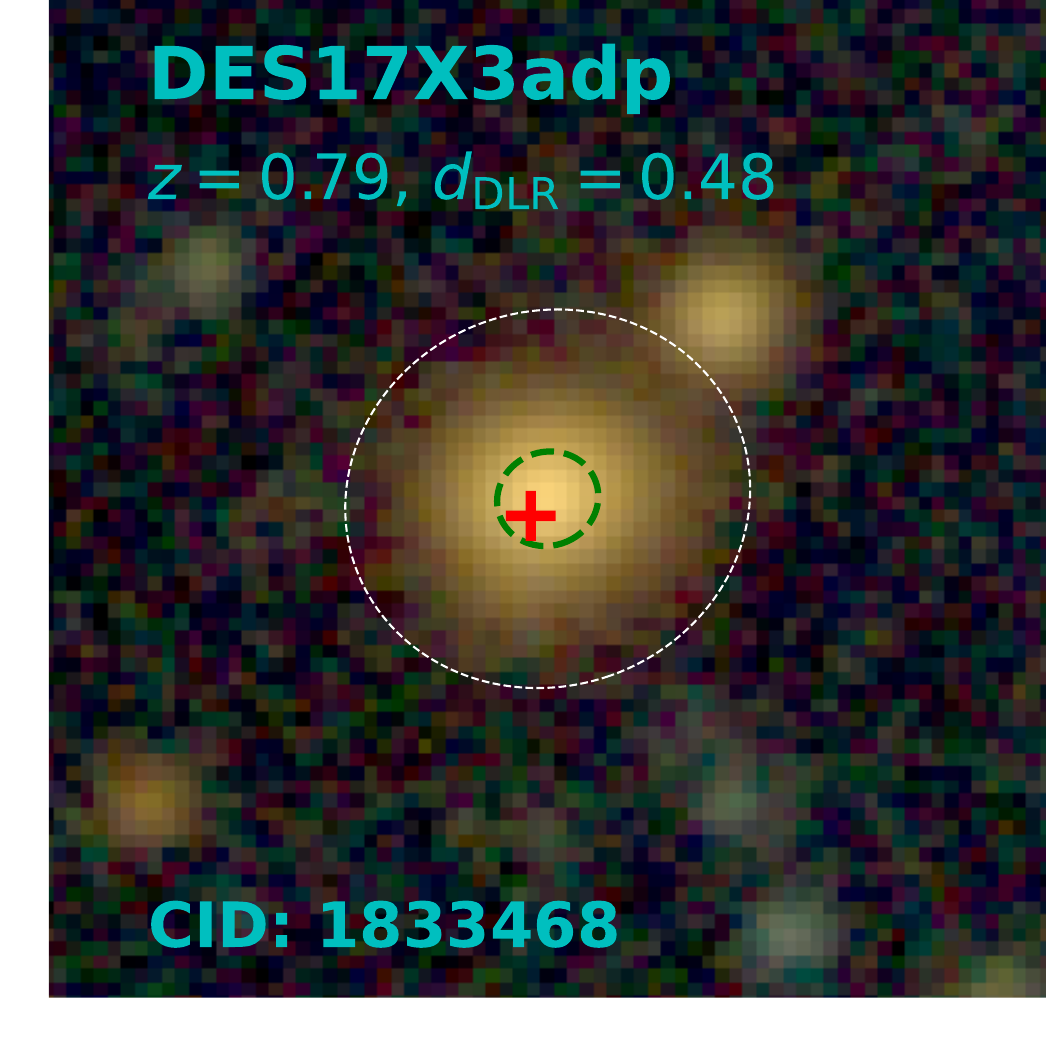}
    \includegraphics[trim={1cm 1cm 0 0},clip, width=0.25 \textwidth]{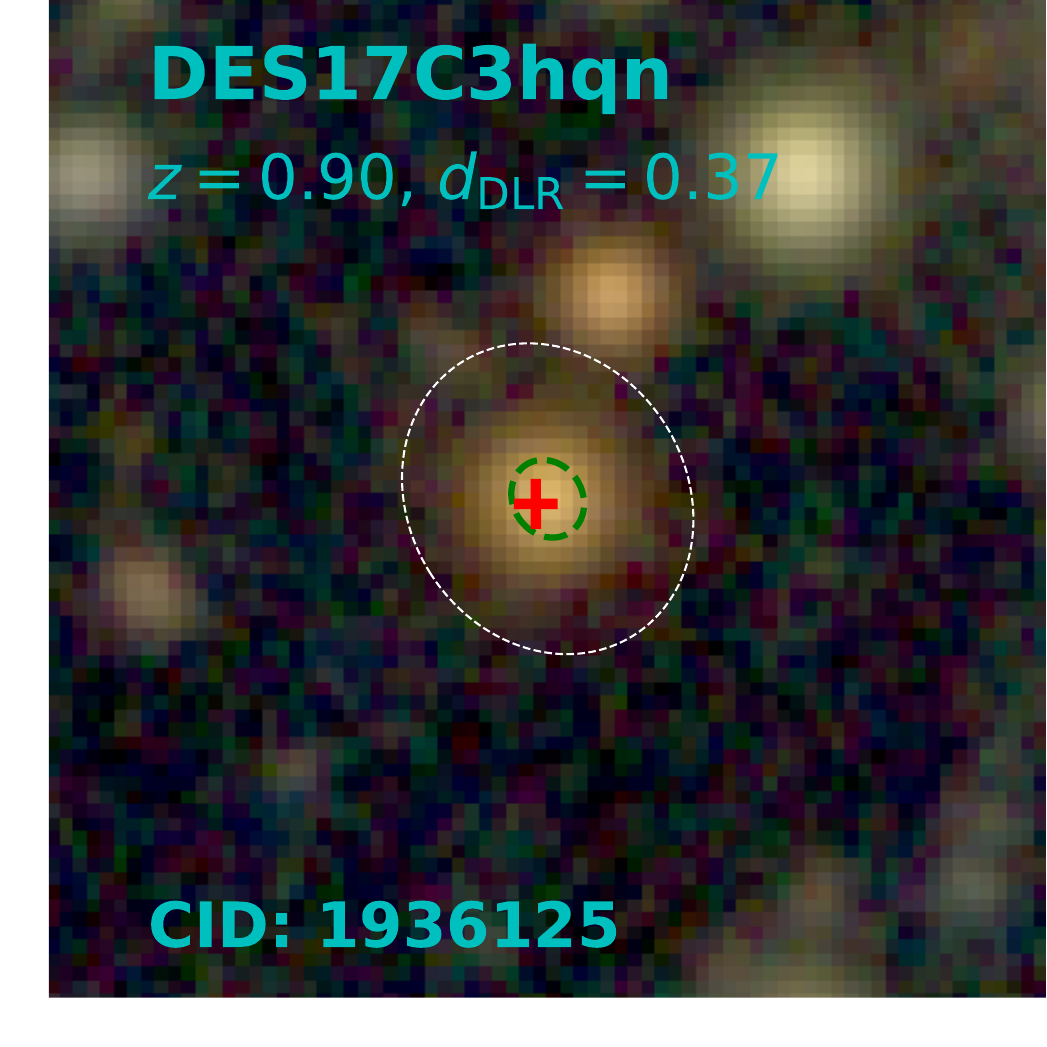}
    \caption{Montage of a selection of DES SN Ia host galaxies with the SN position marked as a red cross, together with the elliptical apertures defined by $\ddlr=1$ (green) and $\ddlr=4$ (white). Both the DES SN name and the DES search pipeline candidate ID (CID) are shown. Each image is 20 arcseconds on each side.}
    \label{fig:montage}
\end{figure*}

\section{Data}
\label{sec:data}

In this section we introduce the dataset that we use in this paper: the \lq five-year\rq\ SN Ia sample \citep{2022MNRAS.514.5159M,2024arXiv240605046S} from the DES SN programme (DES-SN5YR). We use 1533 SNe Ia from the \lq cosmological sample\rq\ of \citet{2024arXiv240102945V} selected from all DES-SN5YR SNe Ia based on the quality of the SN light curves and their light-curve fits (see their table 4). All events are photometrically-identified as SNe Ia \citep{2020MNRAS.491.4277M} with $P_\mathrm{Ia}>0.5$, where $P_\mathrm{Ia}$ is the probability of the event being a SN Ia. All the SNe Ia have a spectroscopic redshift of their host galaxy typically from the follow-up OzDES survey \citep{2020MNRAS.496...19L}. This number differs from the 1541 reported in \citet{2024arXiv240102945V} due to an extra requirement in the bias correction stage (Section \ref{sec:distances}).

Contamination in our SN Ia sample from non-SNIa events has been extensively considered and modelled in previous articles. The DES approach used an improved set of spectrophotomteric core collapse SN templates \citep{2019MNRAS.489.5802V}, with expected level of contamination in cosmological samples studied in \citep{vincenzi2021} and \citep{2023MNRAS.518.1106V}. The level of contamination based on these simulations is expected to be 0.8--3.2 per cent. On the real DES data, different classification algorithms also lead to variations in the SN Ia numbers at a similar level \citep[fig.~11 in][]{2024arXiv240102945V}.

The host galaxies were identified from a catalogue generated from deep, stacked $griz$ images from DES free of SN light \citep{2020MNRAS.495.4040W}, using the Source Extractor software \citep{1996A&AS..117..393B} to measure host positions, structural parameters, and photometry using \citet{1980ApJS...43..305K}-like apertures. SNe were matched to candidate host galaxies using the directional light radius (DLR) method \citep{2006ApJ...648..868S,2016AJ....152..154G,2018PASP..130f4002S}. The host galaxy photometry has been fit with a series of galaxy spectral energy distribution (SED) templates to estimate host stellar masses ($M_*$), SFRs and $U-R$ rest-frame colours following \citet{2006ApJ...648..868S} and \citet{2021MNRAS.501.4861K} as implemented in DES-SN5YR \citep{2020MNRAS.494.4426S}. Galaxy rest-frame colours are reported in \citet{1990PASP..102.1181B} passbands and in the Vega magnitude system \citep{1953ApJ...117..313J}. These rest-frame colours are calculated using the method outlined in \citet{2021MNRAS.501.4861K}, where the best-fitting host galaxy SED for each SN is adjusted with a wavelength-dependent multiplicative function such that the SED exactly reproduces the observed DES $griz$ photometry \citep[often referred to as \lq mangling\rq;][]{2007ApJ...663.1187H, 2008ApJ...681..482C}.

\subsection{SN Ia distances and bias corrections}
\label{sec:distances}

The SN Ia light curves have been fit with the SALT3 light-curve model \citep{2007A&A...466...11G,2021ApJ...923..265K,2023MNRAS.520.5209T} to estimate various light curve parameters. Of particular relevance are $x_1$, a \lq stretch\rq-like parameter \citep{1997ApJ...483..565P} that measures the width of each SN light curve relative to a template, and $c$, a colour parameter that is similar to the rest-frame $B-V$ colour of a SN. Distances to the SNe, $\mu_{\mathrm{obs}}$, are then estimated using \citep[e.g.,][]{1998A&A...331..815T,2006A&A...447...31A}
\begin{equation}
    \mu_{\mathrm{obs}}=m_x + \alpha x_1 - \beta c - \text{M} - \mu_{\mathrm{bias}},
    \label{eqn:mu}
\end{equation}
 where $m_x$, $x_1$ and $c$ are the SALT3 SN light-curve parameters, $\alpha$ and $\beta$ are global parameters parametrizing the stretch--luminosity and colour--luminosity relations, and $\text{M}$ is the absolute magnitude of a SN Ia with $x_1=0$ and $c=0$. $\mu_{\mathrm{bias}}$ is the bias term that corrects for various observational and astrophysical selection effects.

Hubble residuals $\Delta \mu$ are then defined as
\begin{equation}
   \Delta \mu =\mu_{\mathrm{obs}} - \mu_{\mathrm{model}},
    \label{eq:hubbleresiduals}
\end{equation}
where $\mu_{\mathrm{obs}}$ is defined in equation~\ref{eqn:mu} and $\mu_{\mathrm{model}}$ is the distance modulus in a reference cosmology. The Hubble residuals indicate whether the measured distance to a given SN Ia is larger or smaller than that expected in the reference cosmology at the redshift of that SN; negative residuals indicate smaller distances.

%a given SN Ia is brighter or fainter than expected in the reference cosmology; negative residuals indicate a brighter event. 

In our analysis we differ from the main DES-SN5YR analysis in three ways. Firstly, we do not include a \lq $\gamma G_{\mathrm{host}}$\rq\ term in equation~\ref{eqn:mu}. When used, such a term is designed to encapsulate any residual dependencies between the distances estimated to SNe Ia and their host galaxy properties (such as a stellar mass step) that are not corrected for in the light-curve fitting or bias correction steps. As a goal of our paper is to investigate such host galaxy effects, we do not apply this correction.

The second difference is in our use of bias corrections ($\mu_{\mathrm{bias}}$). The bias corrections account for selection biases in the DES SN Ia sample, for example as a function of redshift, SN light curve parameters, or host galaxy properties. The corrections are generally estimated using a Monte Carlo approach modelling the survey detection efficiency and other potential selection effects given assumptions about the underlying SN Ia populations \citep{2009ApJS..185...32K,2010AJ....140..518P,2014A&A...568A..22B}. Early use of simulations modelled distance bias corrections as a function of redshift only \citep{2009ApJS..185...32K,2018ApJ...857...51J} but other approaches have included effects due to stretch and colour \citep{2016ApJ...822L..35S} and incorporated models relating SN properties and their host galaxies \citep{2023ApJ...945...84P}.

The DES-SN5YR analysis uses the approach of \citet{2021ApJ...913...49P}, with bias corrections modelled as a function of $z$, $x_1$, $c$, and $\log(M_*)$ using the \lq Beams with Bias Corrections\rq\ method \citep[BBC;][]{2017ApJ...836...56K}. We refer to this approach as \lq BBC 4D\rq\ \citep[see discussion in][]{2024arXiv240102945V}. This framework includes an underlying model that makes assumptions about how SNe Ia and dust vary by host properties, which is part of the motivation for this paper. Therefore, we use as our main analysis a simpler \lq 1D\rq\ approach, which implements bias corrections only as a function of redshift, accounting for Malmquist-like biases \citep{1922MeLuF.100....1M,1925MeLuF.106....1M} but making no reference to host galaxies in the bias corrections. However, we also include a comparison with the BBC 4D results where relevant. In the BBC framework, SNe that do not achieve a valid bias correction are removed from the analysis. Throughout this work we use the sample of SNe that pass both the 1D and the 4D bias corrections, leaving us with 1533 SNe.

The third difference is that the main DES-SN5YR analysis uses a low-$z$ sample in order to anchor the Hubble diagram, which is important when measuring cosmological parameters. Here, we investigate only the DES-SN5YR SNe Ia and do not include a low-$z$ sample (and hence, we use a reference cosmology for all Hubble residual calculations).

In estimating our distances using BBC 1D, we recover  parameter values of $\alpha=0.159\pm 0.009$ and $\beta=2.73\pm0.05$. These differ from those reported in table 5 of \citet{2024arXiv240102945V}, as those values were estimated using the BBC 4D bias corrections, but are similar to other estimates of $\alpha$ and $\beta$ found for the DES-SN5YR sample using 1D corrections \citep[e.g.,][]{2023MNRAS.519.3046K}.

\subsection{SN--host separations}

SN celestial coordinates ($x_\mathrm{SN}$, $y_\mathrm{SN}$) were measured by DES using point-spread function (PSF) fitting photometry \citep{2015AJ....150..172K}, with the coordinates computed using the weighted-average from each measurement during an observing \lq season\rq. Host galaxies were detected and measured in the DES image stacks by \citep{2020MNRAS.495.4040W} using Source Extractor. We use the celestial coordinates (image centroids) of the Source Extractor detections ($x_\mathrm{host}$, $y_\mathrm{host}$)  measured from images with the same world coordinate system as the images on which $x_\mathrm{SN}$, $y_\mathrm{SN}$ are determined. The statistical uncertainties in both ($x_\mathrm{SN}$, $y_\mathrm{SN}$) and ($x_\mathrm{host}$, $y_\mathrm{host}$) are small: the statistical uncertainties from the PSF fitting of ($x_\mathrm{SN}$, $y_\mathrm{SN}$) and the statistical uncertainties reported by Source Extractor for ($x_\mathrm{host}$, $y_\mathrm{host}$) have a median uncertainty of 8\,mas or 0.03\,pixels. The overall DECam astrometric solution is accurate to 3--6 milliarcseconds or $\simeq$0.02\,pixels \citep{2017PASP..129g4503B}; it is critical that this number is small as the SN and host positions are not measured from the same images and thus represents the systematic uncertainty. All these positional uncertainties are very small and we neglect them in the remainder of our analysis.

These coordinates then define the angular separation (or projected galactocentric distance), $\Delta\theta$, between the SN and the centre of the host galaxy, i.e.,
\begin{equation}
    \Delta\theta = \sqrt{(x_\mathrm{SN}-x_\mathrm{host})^2+(y_\mathrm{SN}-y_\mathrm{host})^2}.
\end{equation}

The DES host galaxies cover a broad range of sizes, stellar masses, inclinations and redshifts; even if converted to physical units (e.g., kpc), the galactocentric distances are not fairly comparable between events if the apparent size of the host galaxy itself varies. Normalising the galactocentric distances using an effective radius or scale length does help, but does not account for inclination effects. We therefore use the DLR as our normalisation of the galactocentric distances, which accounts for the apparent size of a host galaxy in the direction of a SN.

To measure DLRs, we use the Source Extractor basic shape parameters \texttt{A} (semi-major axis) and \texttt{B} (semi-minor axis) that define the elliptical shape of the deblended galaxy. The scales of \texttt{A} and \texttt{B} are set by the second-order moments (variance) of the galaxy's profile along the A and B axes \citep[e.g.,][]{1980SPIE..264..208S}. In particular, we do not use isophotal apertures, where if the same galaxy were observed at increasing redshift, the DLR would become smaller due to cosmological surface brightness dimming \citep[see discussion in, e.g.,][]{1987A&A...183..177I}. The Kron-like apertures used here contain a consistent 94 per cent of the galaxy light \citep{1996A&AS..117..393B}.

The DLR is then defined as\footnote{The definition in \citet{2006ApJ...648..868S} is in terms of Source Extractor \texttt{CXX}, \texttt{CYY}, and \texttt{CXY} ellipse parameters, but is mathematically equivalent.}
\begin{equation}
\mathrm{DLR}=\frac{AB}{\sqrt{(A\sin \phi)^2+(B\cos \phi)^2}},
\end{equation}
where $\phi$ is the angle between the semi-major axis and the vector of interest, in our case a vector connecting the centre of the galaxy to the SN position. The DLR host-matching algorithm chooses the galaxy with the smallest $d_\mathrm{DLR}$, which is defined as the ratio between $\Delta\theta$ and the DLR in the direction of the SN. In other words, $d_\mathrm{DLR}$ measurements are SN--host separations normalised to the size of the host galaxy being compared in the direction to the SN. An upper limit of $\ddlr=4$ is used; \citet{2024ApJ...964..134Q} show that $<2$ per cent of our SNe are likely to be associated with the wrong host galaxy. We show a montage of representative host galaxies and their SN positions in Fig.~\ref{fig:montage}, with the $\ddlr=1$ and $\ddlr=4$ elliptical apertures overplotted. The \ddlr\ values for all DES-SN5YR SNe Ia can be found in the \citet{2024arXiv240605046S} data release.

\begin{figure*}
    \centering
	\includegraphics[width=\columnwidth]{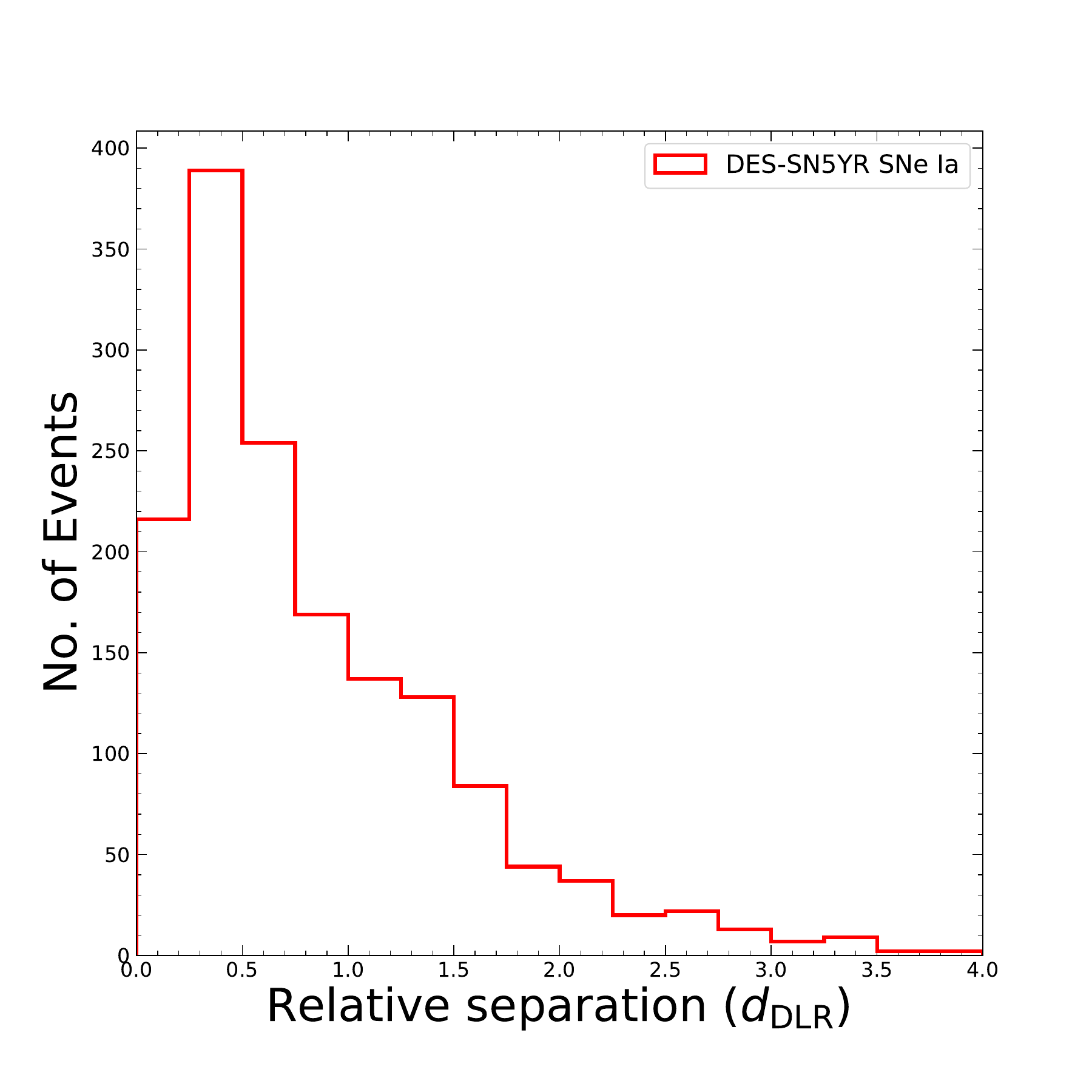} 
    \includegraphics[width=\columnwidth]{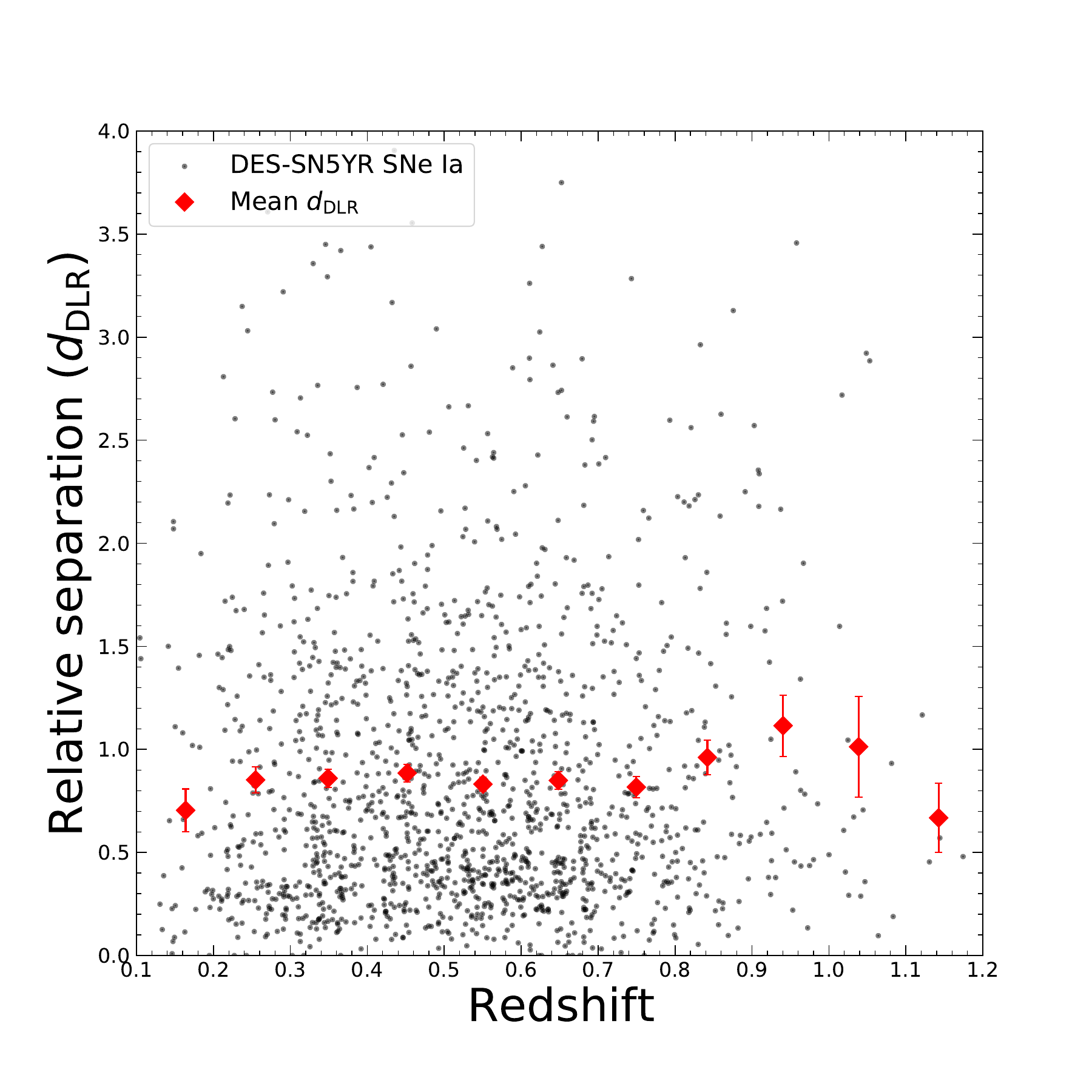}
    \caption{The distribution of \ddlr\ in our sample (left) together with its variation with redshift (right). On the right, the red points show the mean \ddlr\ and uncertainty in bins of redshift.}
    \label{fig:ddlr-dist}
\end{figure*}

The distribution of \ddlr\ in our sample is shown in Fig.~\ref{fig:ddlr-dist}, together with the variation of \ddlr\ with redshift \citep[see also][]{2024ApJ...964..134Q}. We see no redshift-dependent trends.

Seeing effects are important in our host galaxy sample: the number of resolution elements for each of our host galaxies is small given the typical image quality (PSF FWHM) in our image stacks of $\simeq1.3$\arcsec\ \citep{2021MNRAS.501.4861K}. A useful characteristic of $d_\mathrm{DLR}$ is that it is purely empirical and reproducible. There is no model dependency, i.e., fitting a \citet{1963BAAA....6...41S} profile or sophisticated bulge/disc decomposition is not required. This is important in our high-redshift sample where we have very little (if any) morphological information and complicated model fits can become unconstrained.

Because of this convolution with an approximately Gaussian seeing, a useful rule of thumb for physically interpreting DLR is that, for a galaxy with a convolved 2D Gaussian profile, the elliptical shape defining the galaxy detection of radius DLR (on average around $6~\text{kpc}$ for our sample) for any $\phi$ will contain $\simeq68$ per cent of the galaxy light, and  the commonly-used galaxy effective radius (or half light radius) is $\simeq0.67$\,DLR.

\section{SN Ia Properties as a function of galactocentric distance}
\label{sec:SNprop-GCD}

\begin{figure*}
    \centering
	\includegraphics[width=\columnwidth]{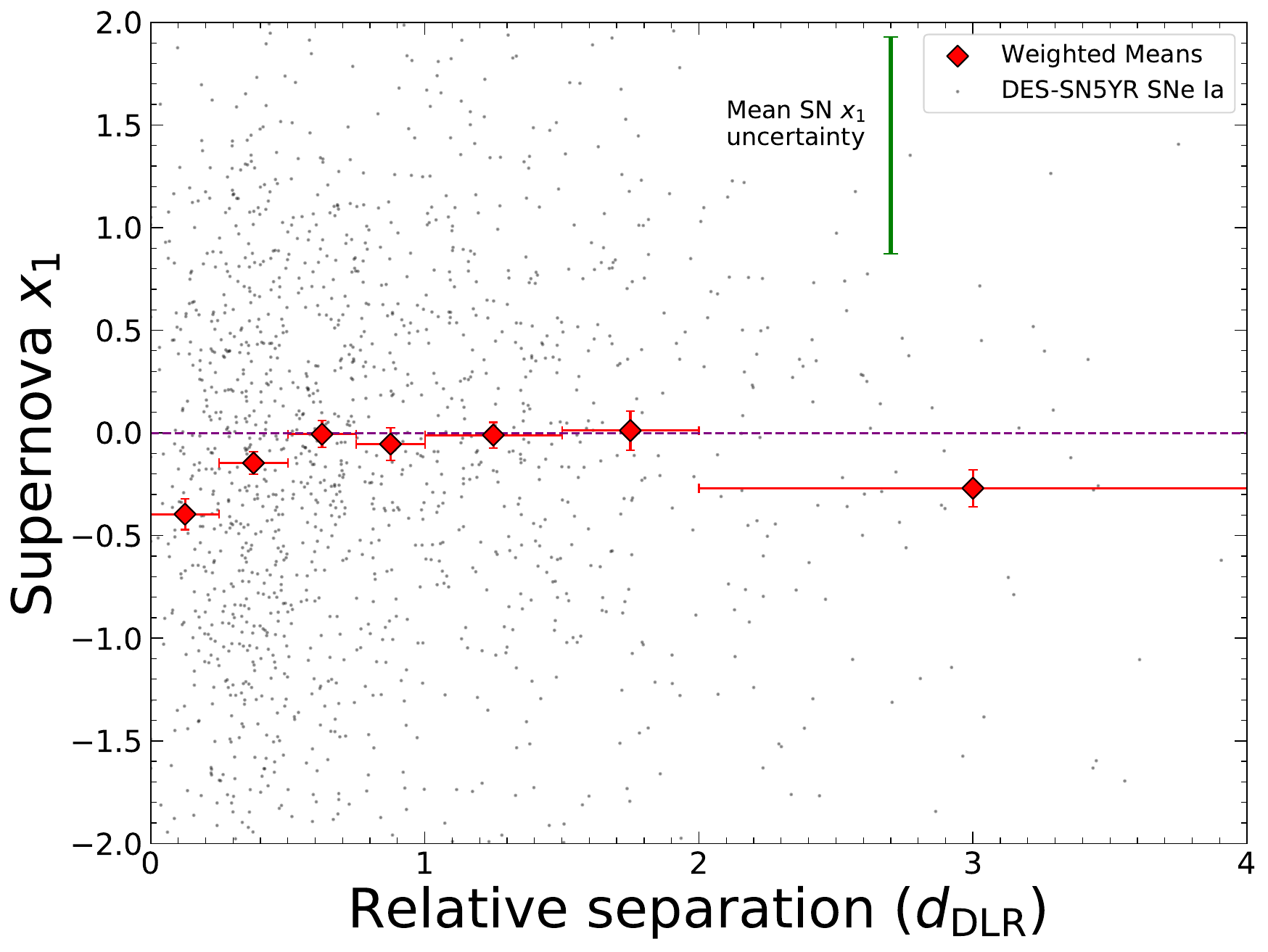} 	
    \includegraphics[width=\columnwidth]{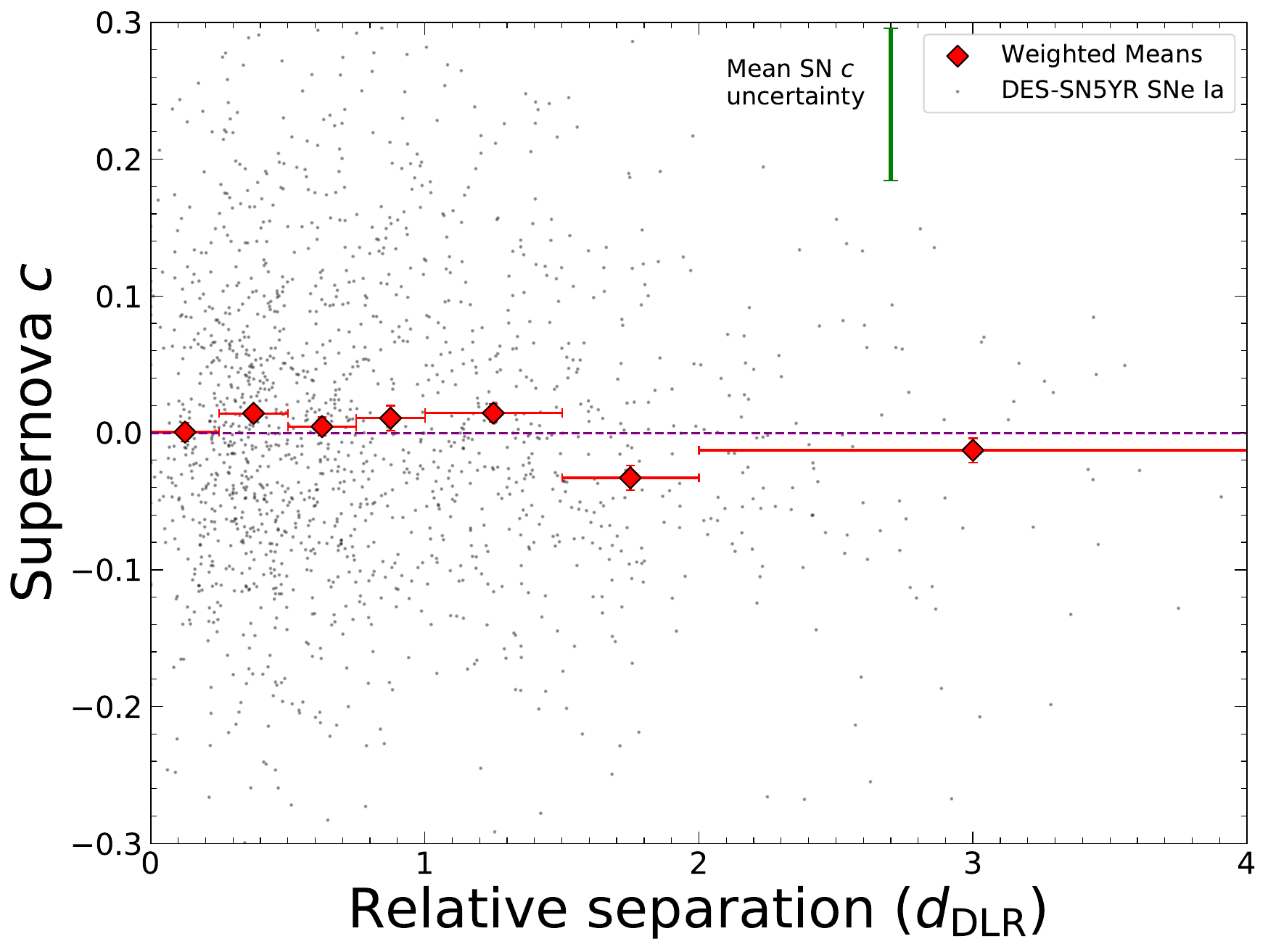}
    \caption{SALT3 SN Ia $x_{1}$ (left) and SN Ia colour $c$ (right) for our sample as a function of \ddlr. The black points are our full sample, and the red points show weighted mean values in bins of \ddlr. The typical uncertainty in $x_1$ and $c$ are shown as the green error bars. No bias corrections have been applied to the $x_1$ and $c$ values. Note the decrease in mean $x_1$ in the centres of host galaxies.}
    \label{fig:x1ddrlC5}
\end{figure*}

In this section we examine how SN Ia properties depend on their galactocentric distance. We begin with the simple photometric properties of light-curve width and colour, and then investigate their Hubble residuals.

\subsection{Light curve properties}
\label{sec:lightcurveprops}

The SALT3 SN $x_{1}$ and $c$ distributions as a function of \ddlr\ are shown in Fig.~\ref{fig:x1ddrlC5}. We give the mean values of the SALT3 $x_{1}$ and $c$ parameters for both inner and outer regions, using different \ddlr\ values to define these, in Table~\ref{tab:meanx1cddlr}. As expected, on average SNe Ia located within the innermost regions ($\ddlr\leq0.5$) of the host galaxy have a smaller $x_1$ than those located at $\ddlr>0.5$ \citep{2023MNRAS.526.5292T}. This means that, on average, SNe within the centres of galaxies are faster evolving than SNe at higher galactic radii. This not surprising: these are likely to be the oldest stellar populations in galaxies, and there are known correlations between light-curve shape and stellar population age \citep[e.g.,][]{2000AJ....120.1479H}.

In SN colour there are no strong trends, but SNe Ia at $\ddlr\leq1$ are slightly redder (higher $c$) than those at $\ddlr>1$, and at $\ddlr>1.5$ the SNe appear bluer on average than those closer to the galaxy centres with a smaller scatter in $c$. There are no red SNe Ia ($c>0.2$) at high \ddlr, consistent with the earlier results of \citet{2018MNRAS.481.2766H} and \citet{2024arXiv240602072G}. As redder and faster SNe Ia are fainter, it is difficult to imagine a selection effect that would bias in favour of these events in the bright inner regions of galaxies.

\begin{table*}
\caption{Weighted mean values of SN Ia light curve properties $x_1$ and $c$ and Hubble residuals in different \ddlr\ ranges.}
\centering
\begin{tabular}{ccccc}
\hline
\ddlr\ range & Number & $\Bar{x_1}$ & $\Bar{c}$ & Mean $\Delta\mu$ \\ 
&of SNe Ia&&&\\
\hline
\multicolumn{1}{|c|}{$\ddlr\leq 0.5$} & 607  & $-0.24\pm0.04$  & $\phantom{-}0.009\pm 0.004$ &$-0.012 \pm 0.010$\\
\multicolumn{1}{|c|}{$\ddlr > 0.5$}   & 926  &  $-0.05\pm0.03$ & $\phantom{-}0.001\pm 0.003$ & $\phantom{-}0.008\pm 0.007$\\
\hline
\multicolumn{1}{|c|}{$\ddlr\leq 1.0$} & 1028 & $-0.16\pm0.03$  & $\phantom{-}0.008\pm 0.003$ & $\phantom{-}0.001\pm 0.007$\\
\multicolumn{1}{|c|}{$\ddlr > 1.0$}   & 505  &  $-0.07\pm0.05$ &$-0.002\pm 0.005$ & $-0.002 \pm 0.009$\\
\hline
\multicolumn{1}{|c|}{$\ddlr\leq 1.5$} & 1294 & $-0.12\pm0.03$  & $\phantom{-}0.010\pm 0.003$ &$\phantom{-}0.002\pm0.006$\\
\multicolumn{1}{|c|}{$\ddlr > 1.5$}   & 239  &  $-0.14\pm0.07$ &$-0.023\pm 0.006$ &$-0.009\pm0.014$\\
\hline
\end{tabular}
\label{tab:meanx1cddlr}
\end{table*}

\subsection{Hubble residuals}
\label{sec:hubbleresiduals}

\begin{figure*}
    \centering
	\includegraphics[width=\columnwidth]{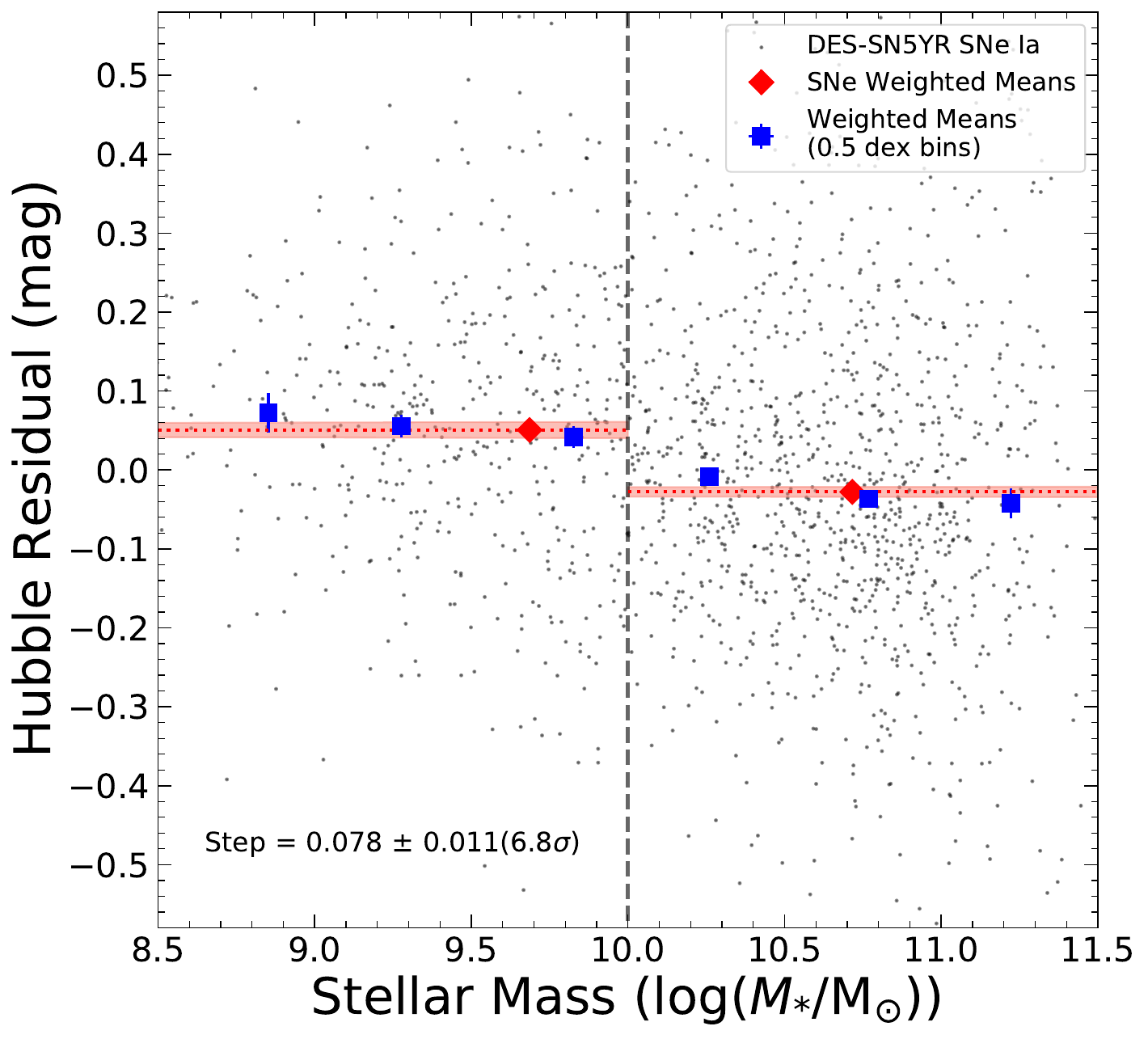}
	\includegraphics[width=\columnwidth]{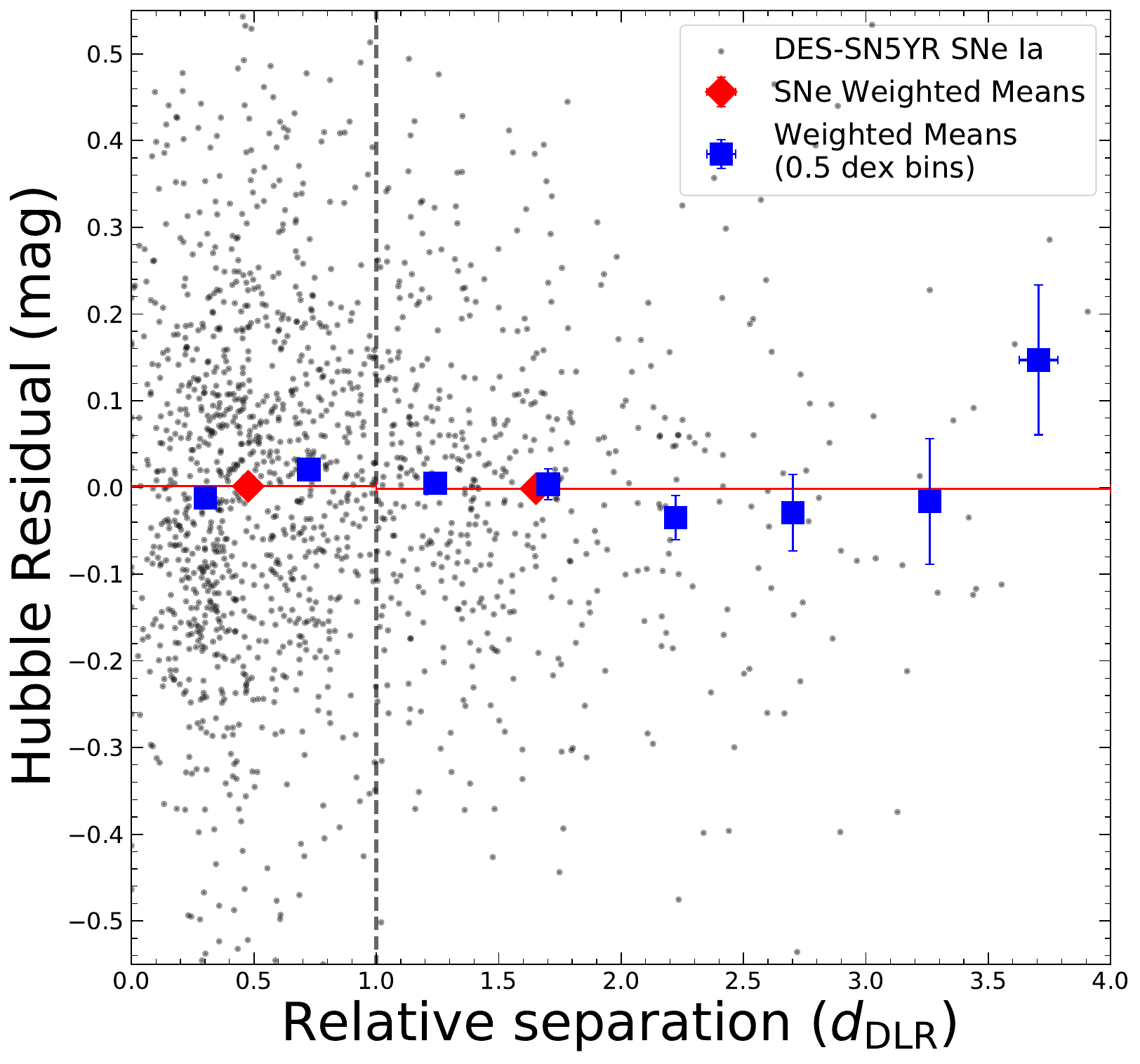}
    \caption{Hubble residuals (equation~\ref{eq:hubbleresiduals}) for the DES-SN5YR photometric sample, with 1D bias corrections (Section~\ref{sec:distances}), as a function of their host galaxy stellar mass (left) and galactocentric distance \ddlr\ (right). In both panels, the blue points show the mean residuals in bins of 0.5\,dex, while the red points and shaded areas show the mean residuals and their uncertainty for SNe Ia to the left and right of the vertical dashed lines. For stellar mass, when divided into bins above and below a stellar mass of $10^{10}\,\msolar$, there is a difference in Hubble residuals of $0.072\pm0.012$\,mag; the so-called \lq mass step\rq. For \ddlr, there is no difference in Hubble residual when splitting the sample at $\ddlr = 1$ ($0.0028\pm0.012$\,mag). However, when restricting the sample to $\ddlr\le1$, and splitting at $\ddlr=0.5$, there is a small step of $0.033 \pm 0.011$\,mag.}
    \label{fig:fullsample_mass_ddlr_step}
\end{figure*}

We next examine the SN Ia Hubble residuals, shown versus host galaxy stellar mass in Fig.~\ref{fig:fullsample_mass_ddlr_step}. We remind the reader that we are using Hubble residuals corrected using 1D bias corrections (i.e., correcting for redshift biases only) as opposed to the nominal values presented in \citet{2024arXiv240102945V} that use the BBC 4D method. This is because we are aiming to understand the causes of Hubble residual--host galaxy correlations, which the 4D approach attempts to model.

As expected based on previous studies of the DES data \citep[e.g.,][]{2023MNRAS.519.3046K}, the SN Ia luminosity step in host galaxy stellar mass (\lq mass step\rq) is clearly seen in the DES-SN5YR sample. Following earlier studies \citep{2010MNRAS.406..782S,2019ApJ...874..150B,2024arXiv240102945V}, we divide the sample at a host galaxy stellar mass of $10^{10}\,\msolar$ and measure a step size of $0.072\pm0.012$\,mag ($\sim6$\,$\sigma$). We make no attempt to optimise the division point of $10^{10}\,\msolar$.% The size of this step is consistent with earlier studies.

\subsubsection{Hubble residuals as a function of \ddlr} 
\label{subsubsec:HRvsdDLR}
 
In the right hand panel of Fig.~\ref{fig:fullsample_mass_ddlr_step} we show the variation in Hubble residual with \ddlr. While previous studies with smaller samples have generally shown null results in similar tests \citep[e.g.,][]{2009ApJ...700.1097H,2012ApJ...755..125G}, here we see a mild trend: when splitting the sample at $\ddlr =0.5$, SNe Ia have more negative Hubble residuals in the inner regions of galaxies. The step is small: $0.020\pm0.012$\,mag (Table~\ref{tab:meanx1cddlr}), which increases to $0.033\pm0.011$\,mag when only considering events within $\ddlr\le1$. However, if we split at $\ddlr=1$, we find no difference between the mean Hubble residuals.

The trend itself is not unexpected given the known relationships between Hubble residual and stellar mass (e.g., left panel of Fig.~\ref{fig:fullsample_mass_ddlr_step}): regions within $\ddlr<0.5$ are likely to be {more} similar to the older, passive stellar environments of massive galaxies {where more negative Hubble residuals are also observed}. %, and the outer regions similar to those in younger lower mass galaxies.
A logical next step is to test how the Hubble residual--$\ddlr$ trend relates to the well-known stellar mass and host galaxy colour steps.

\subsubsection{Host stellar mass step and \ddlr}

\begin{figure*}
    \centering
	\includegraphics[width=\columnwidth]{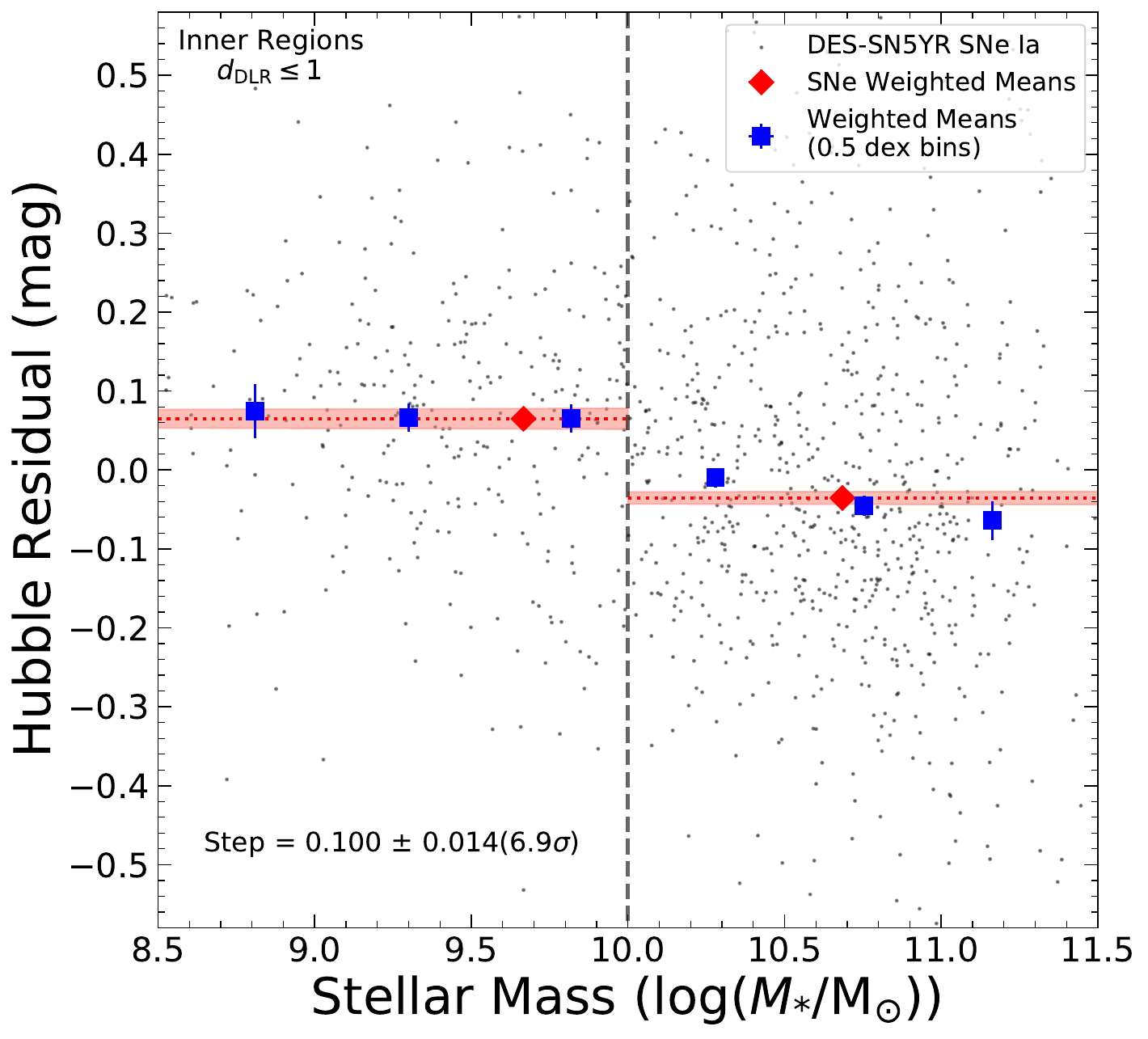}
    \includegraphics[width=\columnwidth]{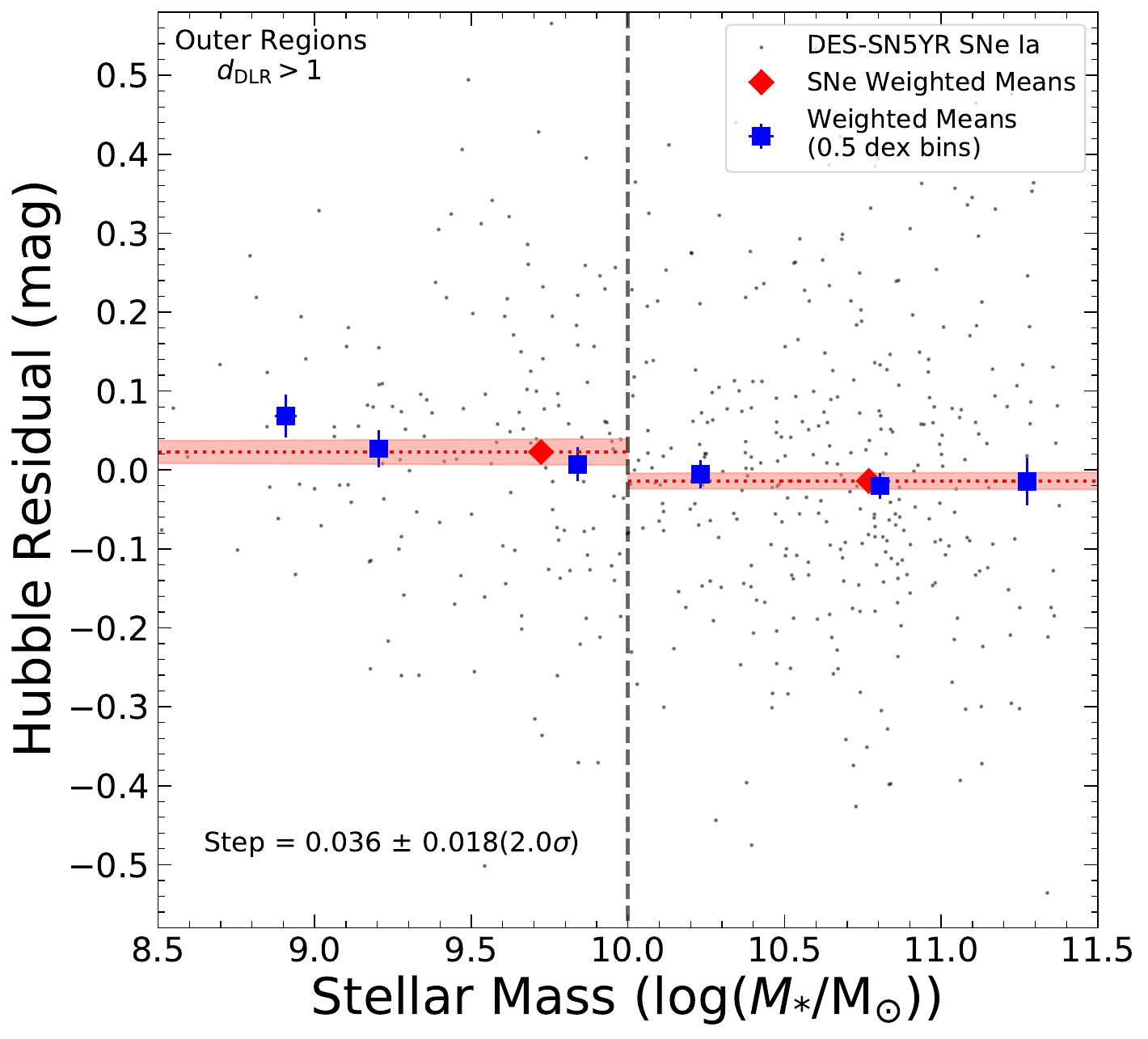}
    \caption{The SN Ia mass step (Fig.~\ref{fig:fullsample_mass_ddlr_step}, left), but limited to SNe Ia with $\ddlr\leq 1$ (left) and $\ddlr>1$ (right). The mass step sizes are $0.100 \pm 0.014$\,mag (left) and $0.036 \pm 0.018$\,mag (right), i.e., the size of the mass step is reduced when restricting to SNe Ia located in the outer regions of their host galaxies.}
    \label{fig:DDLRHMStep}
\end{figure*}

We show the host galaxy stellar mass step separately for SNe Ia in bins of \ddlr\ and find that the size of the stellar mass step does depend on the galactocentric distance (Fig.~\ref{fig:DDLRHMStep}). We split the sample at $\ddlr=1$: within that radius, SNe have a mass step of $0.100 \pm 0.014$\,mag ($6.9\,\sigma$), whereas for $\ddlr>1$ the step is consistent with zero ($0.036\pm0.018$\,mag; $2.0\,\sigma$). The difference in step size between $\ddlr>1$ and $\ddlr\le1$ is $0.064\pm0.023$\,mag. Step sizes and their significance are reported in Table~\ref{tab:hostmasssteps}. At this point, we make no attempt to optimise the step size difference between inner and outer regions, and selected $\ddlr=1$ only for its convenient integer nature. %The greatest disparity occurs at $\ddlr=1$: within that radius, SNe have a mass step of $0.100 \pm 0.014$\,mag ($6.9\,\sigma$), whereas for $\ddlr>1$ the step is consistent with zero ($0.036\pm0.018$\,mag; $2.0\,\sigma$). The difference in step size between $\ddlr>1$ and $\ddlr\le1$ is $0.064\pm0.023$\,mag.}

%Recall that we have made no attempt in our bias correction methodology  to model the host mass step: the step is simply not present in our sample with $\ddlr>1$.

We additionally calculate the root-mean-square deviation (r.m.s), $\sigma_{\mathrm{rms}}$, in the Hubble residuals for SNe Ia in host galaxies on either side of the mass step. The $\sigma_{\mathrm{rms}}$ for outer region SNe Ia is 0.02--0.03\,mag smaller in both lower and higher mass host galaxies than for SNe Ia in the inner regions (Table~\ref{tab:hostmasssteps}). This modest reduction may indicate a more robust standardization for SNe in the outer regions compared to the inner regions, consistent with results in the literature \citep[e.g.,][]{2018MNRAS.481.2766H,2020ApJ...901..143U}.

\begin{table*}
\caption{Statistics related to the steps in SN Ia luminosity as a function of stellar mass (the mass step) in our samples.}
\begin{tabular}{lccccccc}
\hline
%Sample& \begin{tabular}[c]{@{}c@{}}Number \\ (LM hosts, HM hosts)\end{tabular}  & \begin{tabular}[c]{@{}c@{}}Mean $\Delta\mu$ in \\ low-mass hosts\end{tabular} & \begin{tabular}[c]{@{}c@{}}Mean $\Delta\mu$ in \\ high-mass hosts\end{tabular} & \begin{tabular}[c]{@{}c@{}}Size of the \\ mass step\end{tabular} & \begin{tabular}[c]{@{}c@{}}Significance\\of step\end{tabular} & \begin{tabular}[c]{@{}c@{}}r.m.s in \\ low-mass hosts\end{tabular} & \begin{tabular}[c]{@{}c@{}}r.m.s in \\ high-mass hosts\end{tabular}\\
Sample & Number & Mean $\Delta\mu$ in & Mean $\Delta\mu$ in & Size of the & Significance & r.m.s. in&r.m.s. in\\
&(LM hosts, HM hosts)&low-mass hosts&high-mass hosts&mass step&of step&low-mass hosts&high-mass hosts\\
\hline
Full Sample& 1533 (472, 1061)  & $0.051 \pm 0.009$& $-0.028 \pm 0.007$& $0.078 \pm 0.011$& $6.8\sigma$ & 0.199 & 0.226\\
$\ddlr\leq 1$ &  1028 (310, 718) & $0.065 \pm 0.012$& $-0.035 \pm 0.009$& $0.100\pm0.014$& $6.9\sigma$ &0.205 & 0.232\\
$\ddlr>1$ &  505 (162, 343) & $0.023 \pm 0.014$& $-0.014 \pm 0.011$& $0.036 \pm 0.018$& $2.0\sigma$ & 0.180 & 0.212\\
\hline
$z<0.6$ & 982 (329, 653) & $0.049\pm0.009 $ & $-0.034\pm0.008 $ & $0.083\pm0.012 $ & $ 6.9 \sigma $ & 0.166 & 0.201 \\
$\ddlr\leq 1$  & 652 (220, 432) & $0.063\pm0.012 $ & $-0.041\pm0.010 $ & $0.104\pm0.015 $ & $ 6.7 \sigma $ & 0.174 & 0.211\\
$\ddlr>1$  & 330 (109, 221) & $0.021\pm0.014 $ & $-0.022\pm0.012 $ & $0.043\pm0.018 $ & $ 2.4 \sigma $ & 0.141 & 0.178 \\
\hline
$P_\mathrm{Ia}>0.9$ & 1442 (449,991) & $0.050\pm0.009$ & $-0.028\pm0.007$ & $0.079\pm0.012$ & $6.8\sigma$& 0.195& 0.222\\
$\ddlr\leq 1$  & 968 (295, 673)& $0.065\pm0.012$ & $-0.034\pm0.009$ & $0.100\pm0.015$ & $6.8\sigma$ & 0.203 & 0.227 \\
$\ddlr>1$  & 472 (154, 318) & $0.021\pm0.014$ & $-0.018\pm0.012$ & $0.039\pm0.018$ & $2.1\sigma$ & 0.174 & 0.212 \\
\hline
\end{tabular}
\label{tab:hostmasssteps}

\end{table*}

\begin{figure*}
    \centering
	\includegraphics[width=\columnwidth]{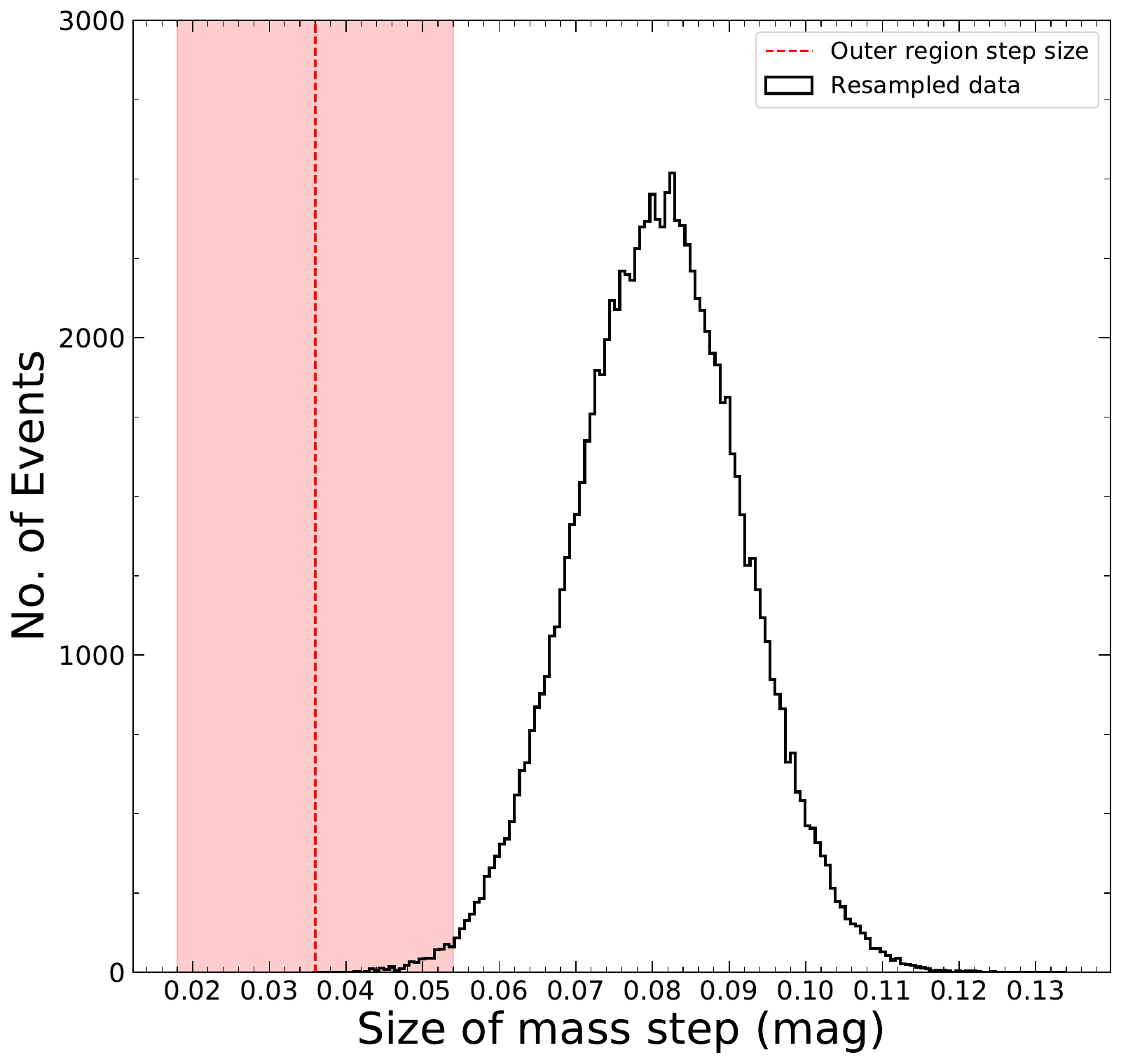}
 	\includegraphics[width=\columnwidth]{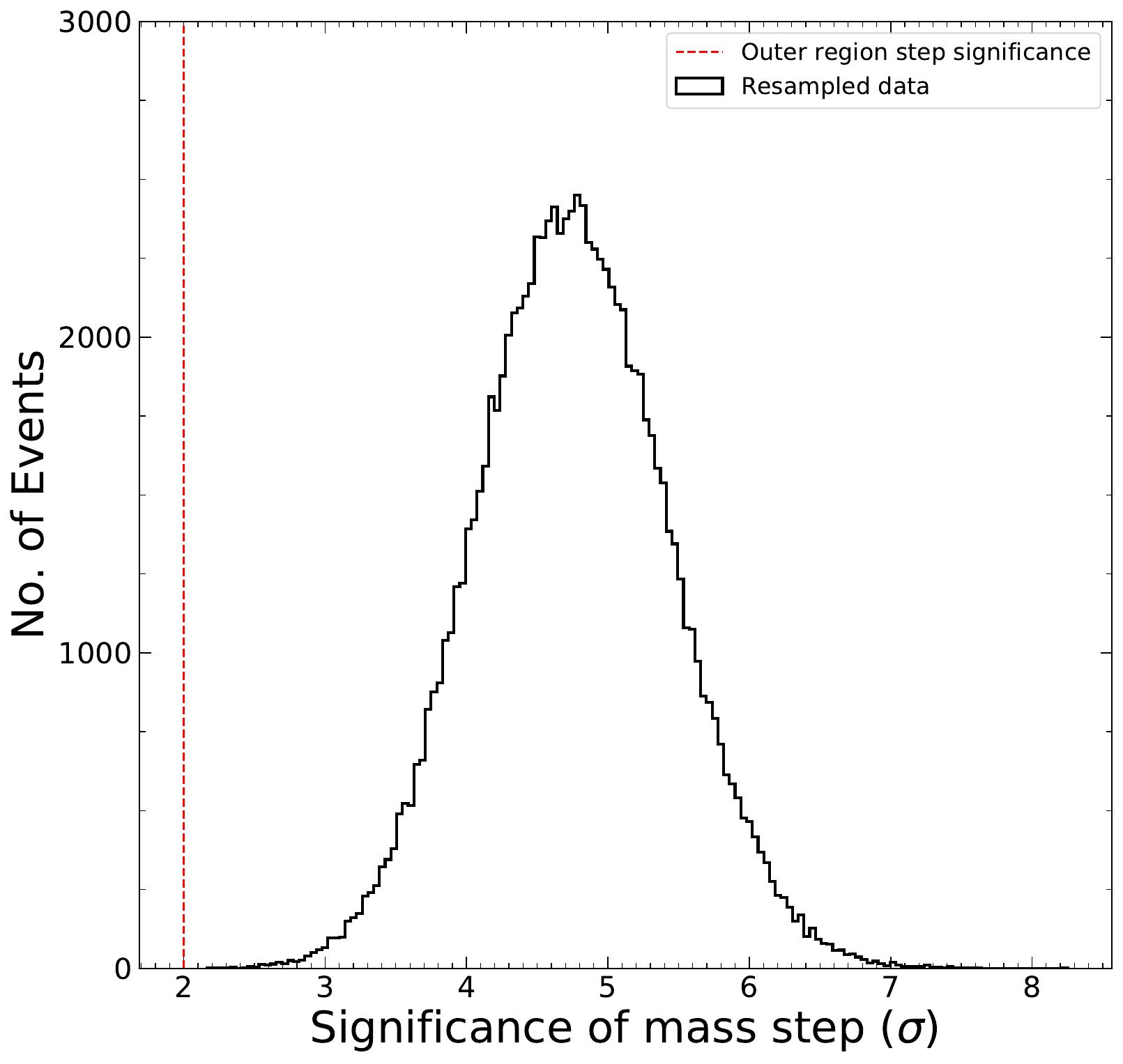}
     \caption{Bootstrap resampling test of the significance of the reduction of the mass step observed in SNe Ia in the outer regions of their host galaxies ($\ddlr\le1$). SNe are selected with replacement from the full sample to match the size and colour distribution of the true sample of SNe Ia in the outer regions. We perform $10^5$ such resamples and show histograms of the results. Left: The distribution of host mass step size in the $10^5$ resamples. The distribution of the mass step in the resampled SNe is centred on a step size of 0.08\,mag, consistent with the host mass step observed in our full sample. Right: The distribution of host mass step significance. On average, a statistically significant step is found for our resampled SNe. The dashed vertical lines show the actual values measured in the real data for SNe in the outer regions.}
    \label{fig:SigOfStep}
\end{figure*}

We verify the robustness of the change in mass step sizes at $\ddlr\le1$ and $\ddlr>1$ by bootstrap resampling. We select (with replacement) the same number of SNe as in the outer regions (505) randomly from the full sample of 1533 events. We weight this selection so that the resampled SNe have the same colour distribution as those truly in the outer regions. We then calculate the size and significance of any host mass step in each resample. The results for $10^5$ draws (resamples) are shown in Fig.~\ref{fig:SigOfStep}. On average, the step size in resamples is consistent with the full sample, although less significant given the smaller sample size ($\Bar{\sigma} = 4.7 \pm 0.75$ versus $\sigma\simeq6$). We conclude that the reduced mass step in SNe Ia located in the outer regions is unlikely to be due to random chance.

We perform two additional tests of the change in mass step: we restrict our sample to events at $z<0.6$ where selection effects are smaller, and we select events with $P_\mathrm{Ia}>0.9$ (instead of 0.5; see Section~\ref{sec:data}) to remove those events least likely to be SNe Ia. The results are in Table~\ref{tab:hostmasssteps}. The difference in step sizes between $\ddlr>1$ and $\ddlr\le1$ for these two samples is $0.061\pm0.023$\,mag in both cases, compared to $0.064\pm0.023$\,mag in our full sample, an almost identical result.

\begin{figure*}
    \centering
    \includegraphics[width=\columnwidth]{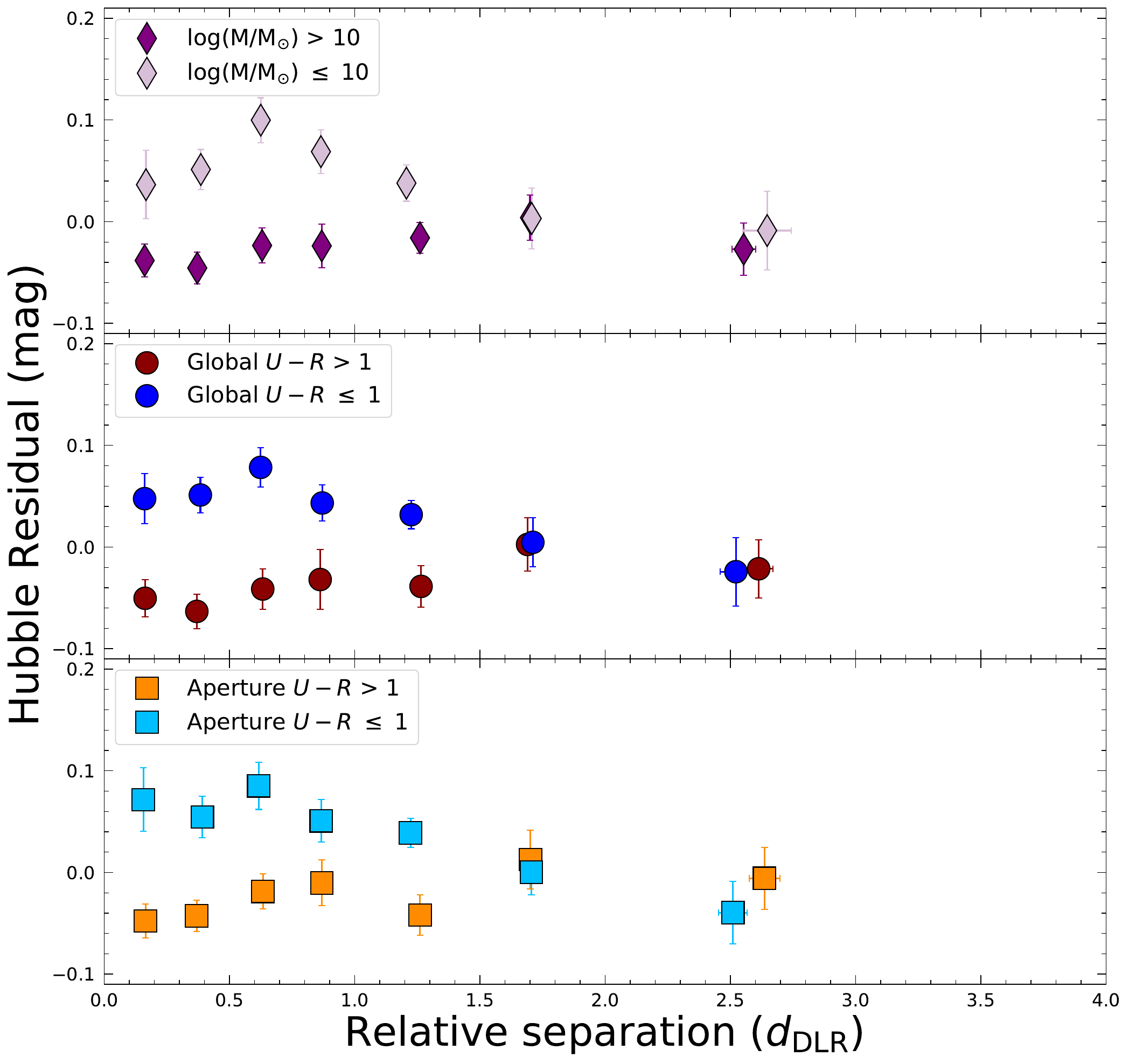}	\includegraphics[width=\columnwidth]{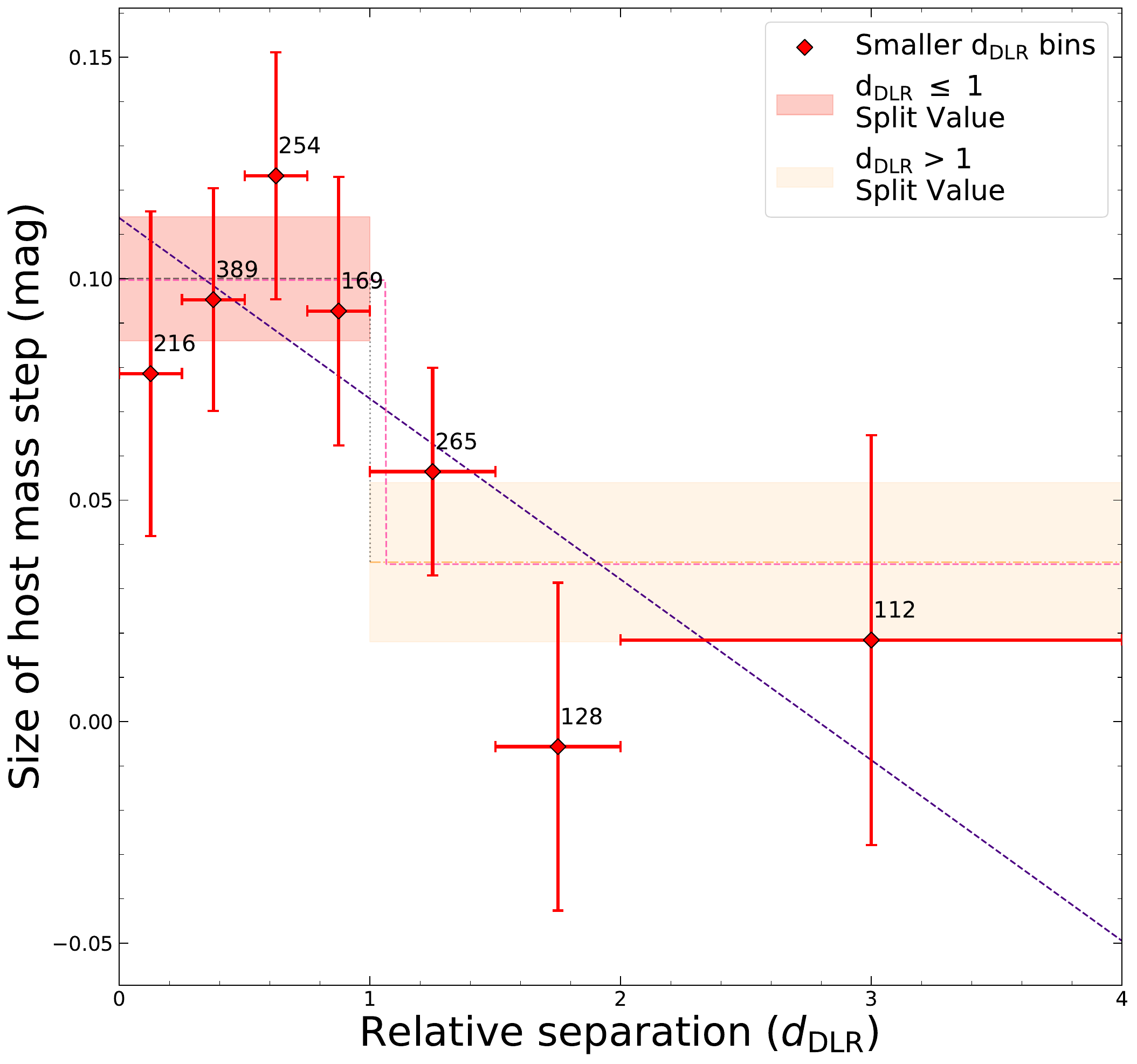}

    \caption{The host mass step observed in smaller \ddlr\ bins of 0.25\,dex over $0\leq\ddlr\leq1$, 0.5\,dex over $1<\ddlr\leq2$, and 2\,dex at $\ddlr>2$. Left: The Hubble residuals split into low and high stellar mass samples (top panel), split by the global host galaxy rest-frame $U-R$ colour (middle panel), and split by an aperture $U-R$ colour (lower panel). The aperture colours are described in Section~\ref{sec:hostcolour-ddld-HR}. Right: The magnitude of the host mass step in these \ddlr\ bins. We fit both a step function and a linear function to our data, and find both have similar goodness-of-fit values. Numbers show the number of SNe Ia within each \ddlr\ bin.}
    \label{fig:BinnedDDLRStep}
\end{figure*}

To determine if the observed difference in mass step size between inner and outer regions have sensitivity to the choice of $\ddlr = 1$, we use smaller \ddlr\ bins, each with a similar number of SNe Ia. We also fit a straight line to the data. The results are shown in Fig.~\ref{fig:BinnedDDLRStep}. The figure shows a clear difference in step size between $\ddlr\le1$ and $\ddlr>1$, i.e., the result in Fig.~\ref{fig:DDLRHMStep} is not dependent on the choice of $\ddlr=1$ to divide the sample.

\begin{figure*}
    \centering
    \includegraphics[width=\columnwidth]{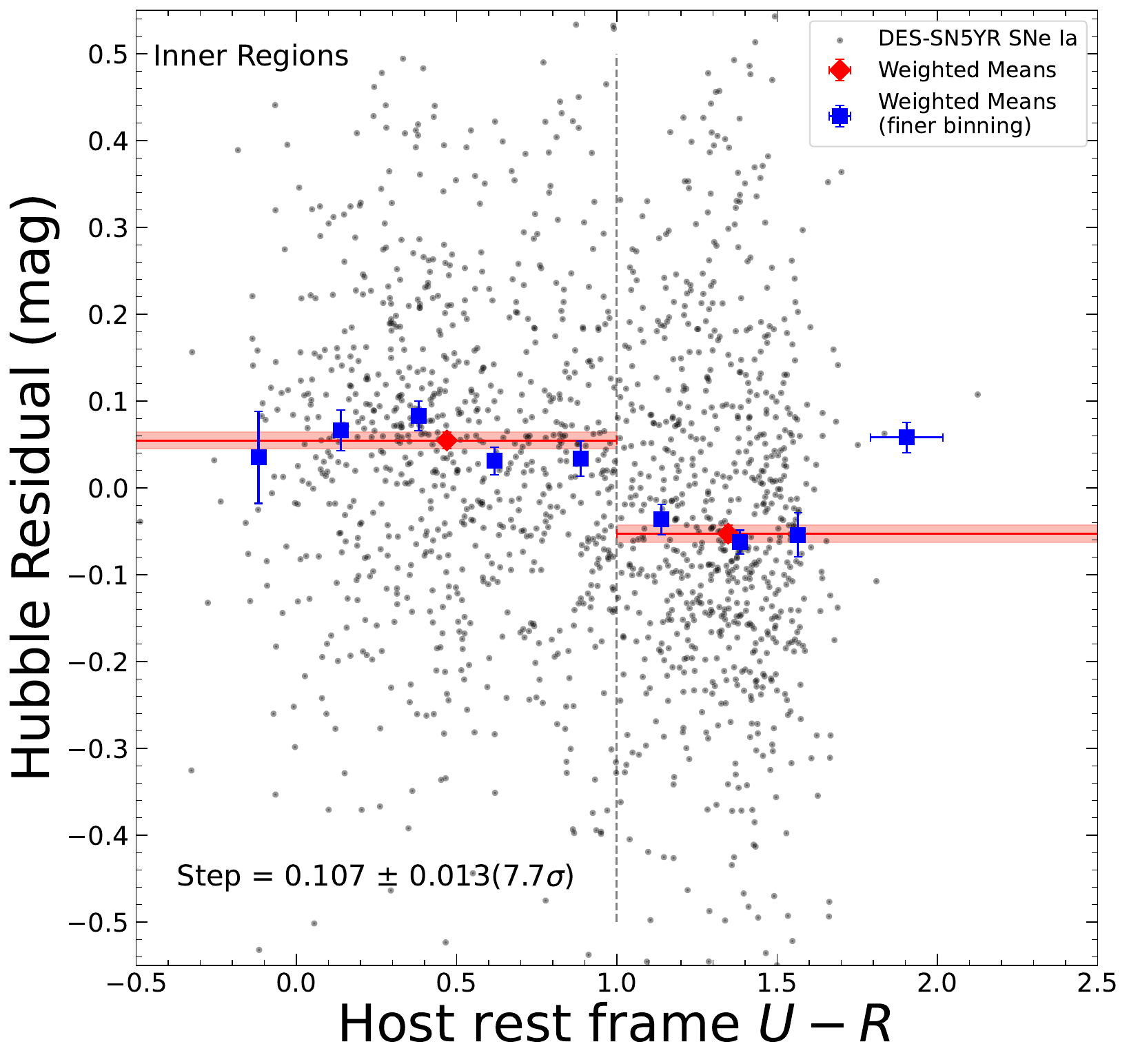}\includegraphics[width=\columnwidth]{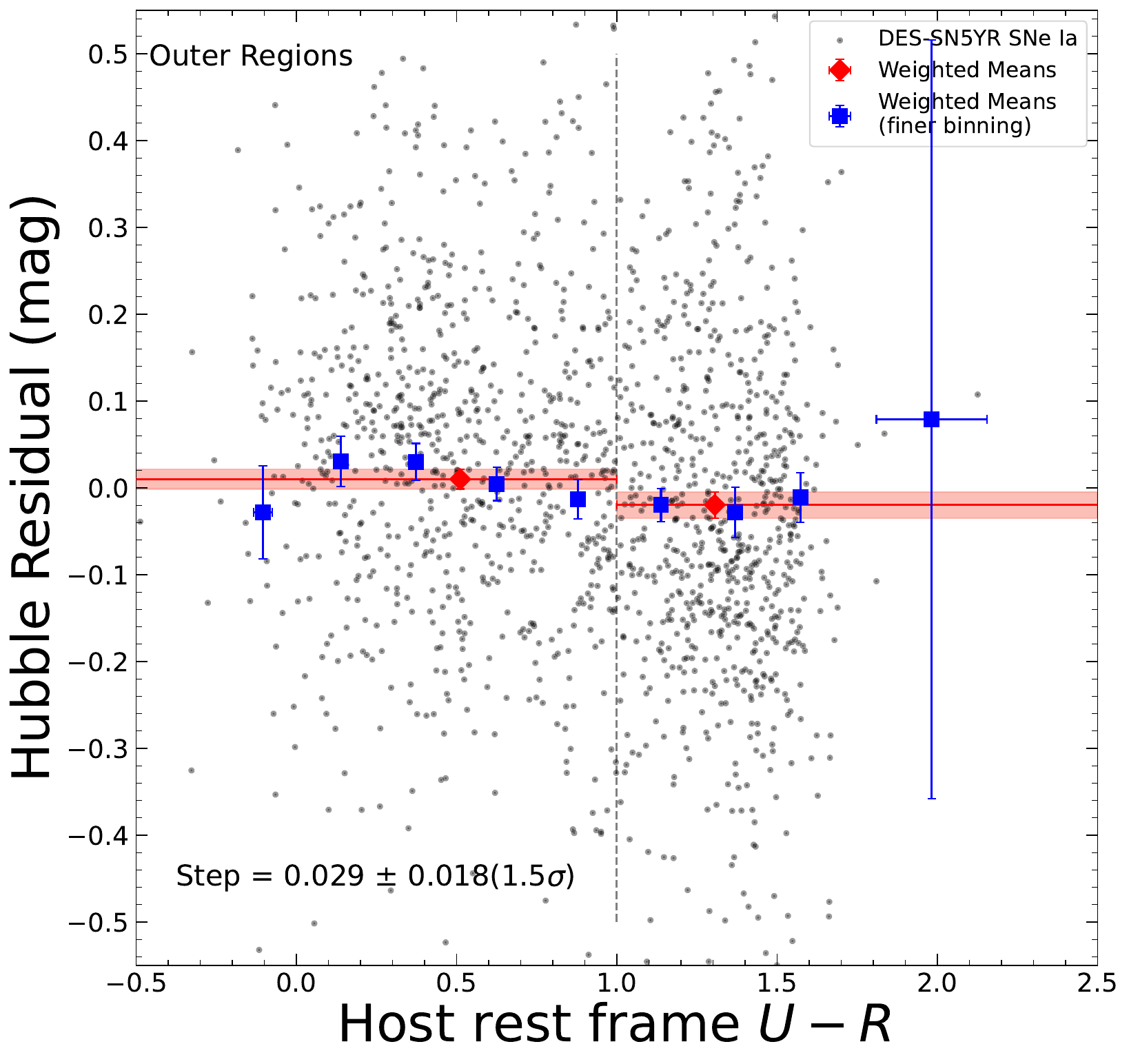}	
    \caption{As Fig.~\ref{fig:DDLRHMStep} but for host galaxy rest-frame $U-R$ colour. The step sizes are $0.107 \pm 0.013$\,mag (left) and $0.029 \pm 0.018$\,mag (right).}

    \label{fig:URsteps}
\end{figure*}

\subsubsection{Host colour steps and \ddlr}
\label{sec:hostcolour-ddld-HR}

Complementary to stellar mass, host galaxy colour also correlates strongly with Hubble residual with at least equal (if not greater) significance than stellar mass \citep{2018A&A...615A..68R,2023MNRAS.519.3046K,2022A&A...657A..22B,2022MNRAS.515.4587W,2023MNRAS.520.6214W}. Galaxy colours trace a combination of the age of the stellar population and to a lesser extent the integrated dust extinction, and the correlations between Hubble residual and galaxy colour are often interpreted as the age of the SN progenitor system (or a combination of the progenitor and its surrounding environment such as dust along the line of sight) as being the driver of the step.

We show the host galaxy rest-frame $U-R$ colour versus Hubble residuals in Fig.~\ref{fig:URsteps}. We split the SN Ia sample at host $U-R=1$ following \citet{2021MNRAS.501.4861K,2023MNRAS.519.3046K}. The step sizes are similar to those for stellar mass: SNe Ia within $\ddlr\le1$ have a strong Hubble residual step between red and blue hosts, while the step between SNe in red and blue hosts at $\ddlr>1$ is reduced to $\sim1.5$\,$\sigma$. 

The global galaxy colour traces the galaxy-integrated properties. To understand better whether the observed \ddlr\ effects are driven by stellar population age, we measure colours for inner and outer regions separately. We follow the method of \citet{2021MNRAS.501.4861K} and \citet{2023MNRAS.519.3046K}, but use an elliptical aperture defined by $\mathrm{DLR}=1$ and an  annulus with $1<\mathrm{DLR}<4$. Splitting on the colour of the region the SN occurred in (i.e., inner colour for SNe with $\ddlr\le1$ and outer colour for SNe with $\ddlr>1$) makes negligible difference to the size of the step compared to the global colour. 

To assess whether the Hubble residual step at $\ddlr=0.5$ is caused by the transition between older and younger stellar populations, we also plot Hubble residual versus \ddlr\ for SNe in blue regions and red regions. We compare this to the trend between the mass step and \ddlr\ in Fig.~\ref{fig:BinnedDDLRStep}.

\subsection{Inclination effects}
\label{sec:inclination}

Our \ddlr\ measurements give a projected normalised galactocentric SN distance, and therefore present a lower limit to the true (deprojected) galactocentric distance. This means that while SNe at high \ddlr\ are likely to truly be located in the outer regions, SNe with lower \ddlr\ can be more ambiguous. If galactocentric distance does influence the SN Ia Hubble residual, this effect may be diluted due to our (unavoidable) use of projected separation.

We investigate this effect by calculating the apparent eccentricity, $e_{\mathrm{gal}}$, of each host galaxy as 
\begin{equation}
    e_{\mathrm{gal}} = \sqrt{1-\frac{B^2}{A^2}}\,.
    \label{eq:eccentricity}
\end{equation}
We then identify a sample with $e_{\mathrm{gal}}<0.5$ that has smaller inclinations (i.e., that are more face on galaxies), and limit our inner region sample to only those host galaxies.

The magnitude of the host mass step in this sample remains consistent at $0.095\pm0.023$ mag, while the significance decreases to $4.0\sigma$ due to the smaller sample size. We conclude that there is no evidence that inclination is affecting the \ddlr\ results.

\section{Discussion}
\label{sec:discussion}

We have uncovered a relationship between the Hubble residuals of SNe Ia and the projected, normalized galactocentric distance of the SN from its host galaxy. After standardization for stretch and colour, SNe Ia in the inner regions of their host galaxies ($\ddlr\le0.5$) have more negative Hubble residuals (i.e., appear brighter) than those at moderate ($0.5<\ddlr\le1$) galactocentric distances (Fig.~\ref{fig:fullsample_mass_ddlr_step}). More significantly, we have found that the magnitude of the step in Hubble residuals that occurs at stellar masses around $10^{10}\,\msolar$ is dependent upon the galactocentric distance: the step is large for SNe Ia with $\ddlr\le1$ but negligible for SNe Ia with $\ddlr>1$ (Fig.~\ref{fig:DDLRHMStep}). This result adds to the growing body of evidence that the size mass step can be reduced by a variety of methods and sample selections \citep{2021ApJ...909...26B,2021ApJ...912...71B}.

Two leading astrophysical explanations for the host mass step are dust along the line of sight to the SN, and SN Ia progenitor age (or a combination of both). Here we discuss the implications of our findings for each model.

\subsection{Dust}
\label{subsec:disc_dust}

Previous studies have proposed an explanation of the host mass step in the form of a change in the ratio of total-to-selective extinction of the dust ($R_V$) in low and high-mass galaxies \citep{2021ApJ...909...26B}. A key motivation for the introduction of the \citet{2021ApJ...909...26B} model is that it can naturally explain the observation that the size of the mass step increases for redder SNe Ia. This trend between mass step size and SN colour is present (before the BBC 4D bias correction) in DES-SN5YR data \citep{2024arXiv240102945V}, as well as the Pantheon+ \citep{2022ApJ...938..110B} and Amalgame \citep{2024MNRAS.529.2100P} compilations. However, \citet{2024arXiv240602072G} found no such significant trend between SN Ia Hubble residual and SN Ia colour in the low-$z$ ZTF sample. 

$R_V$ determines the amount of dimming at a given wavelength for a total amount of reddening. Consider two populations of SNe with different dust screens: the two populations can experience the same distribution of reddening (i.e., the $\tau_E$ for each dust screen is identical) while experiencing different amounts of (for example) rest-frame $B$-band extinction if the average $R_V$ is different. This difference in the extinction experienced in the rest-frame $B$-band ($\Delta m_B$) is quantified by
\begin{equation}
    \Delta m_B = (R_V+1)E(B-V)\,.
    \label{eq:delta_mb_rv}
\end{equation}
If the $R_V$ distributions are different for the two populations, the size of the difference in $m_B$, and thus Hubble residuals through equation~\ref{eqn:mu}, increases with increasingly red SN colours. Note that to explain the step, the extinction $R_V$ must be larger in high-mass galaxies (opposite to trends observed in star-forming galaxies \citep{2018ApJ...859...11S}) and passive galaxies (which follows the trend observed in that paper). Thorough investigations of the relationships between extinction $R_V$ and attenuation $R_V$ in the context of SN Ia cosmology can be found in \citet{2023A&A...680A..56D} and \citet{2024arXiv240605051P}.

The difference between the observed mass step in inner and outer regions could be due to either a complete lack of dust in the outer regions of galaxies (i.e., there are no SNe with sufficiently large $E(B-V)$ to be affected by a difference in $R_V$) and/or a lack of evolution of the dust laws in the outskirts as galaxies transition from low to high mass (i.e, the $R_V$ distribution is consistent in the outskirts of all galaxies). The signature of the former would be a deficiency of red SNe from the overall SN colour distribution in the outer regions compared to the inner regions. The latter would show up when plotting Hubble residual, split between high and low-mass hosts, against SN colour: inner regions would show diverging Hubble residuals for redder SNe, while outer regions would show no difference in the size of step as a function of SN colour. An evolution of the dust law may also present as a difference in the best-fitting $\beta$ values from Eq. \ref{eqn:mu} between the high and low-mass samples in our two regions, because of the contribution of Eq. \ref{eq:delta_mb_rv} to the overall colour--luminosity relation $\beta$. We next investigate these possibilities.

\subsubsection{SN colour distributions}
\label{subsubsec:c_dists}

\begin{table*}
    \caption{Best-fitting parameters for the intrinsic + dust colour model. DES-SN5YR is the nominal sample (no redshift selection, but with $x_1$ and $c$ restricted as in \citet{2024arXiv240102945V}. In the middle section we loosen the $c$ selection from $-0.3<c<0.3$ to $-0.5<c<0.5$, and select SNe Ia with $z<0.6$. The lower section shows the \citet{2024arXiv240602072G} ZTF results.}
    \centering
    \begin{tabular}{|l|c|c|c|}
        \hline
        Selection & $\mu_c$ & $\sigma_c$ & $\tau_E$ \\
        \hline
        DES-SN5YR; nominal & $-0.057 \pm 0.003$ & $0.038 \pm 0.003$ & $0.093 \pm 0.004$ \\
        $d_{\mathrm{DLR}}\le1$ & $-0.056 \pm 0.003$ & $0.038 \pm 0.003$ & $0.100 \pm 0.005$ \\
        $d_{\mathrm{DLR}}>1$ & $-0.056 \pm 0.004$ & $0.034 \pm 0.003$ & $0.093 \pm 0.007$ \\
         $d_{\mathrm{DLR}}\le1.5$ & $-0.054 \pm 0.003$ & $0.037 \pm 0.003$ & $0.100 \pm 0.005$ \\
        $d_{\mathrm{DLR}}>1.5$ & $-0.062 \pm 0.004$ & $0.030 \pm 0.003$ & $0.070 \pm 0.007$ \\
        \hline
        DES-SN5YR; $z<0.6,|c|<0.5$ & $-0.048\pm0.003$ & $0.034\pm0.003$ & $0.127\pm0.006$ \\
        $d_{\mathrm{DLR}}\le1$ & $-0.047\pm0.003$ & $0.036\pm0.003$ & $0.132\pm0.007$ \\
        $d_{\mathrm{DLR}}>1$ & $-0.050\pm0.004$ & $0.035\pm0.004$ & $0.115\pm0.009$ \\
        $d_{\mathrm{DLR}}\le1.5$ & $-0.046\pm0.003$ & $0.037\pm0.003$ & $0.131\pm0.007$ \\
        $d_{\mathrm{DLR}}>1.5$ & $-0.056\pm0.004$ & $0.030\pm0.004$ & $0.093\pm0.009$ \\
        \hline
        ZTF \citet{2024arXiv240602072G} Full sample & $-0.086 \pm 0.004$ & $0.029 \pm 0.005$ & $0.157 \pm 0.007$ \\
        ZTF \citet{2024arXiv240602072G} Dustless sample & $-0.092 \pm 0.008$ & $0.032 \pm 0.008$ & $0.099 \pm 0.011$ \\
        ZTF \citet{2024arXiv240602072G} Non dustless sample& $-0.081 \pm 0.005$ & $0.029 \pm 0.006$ & $0.168 \pm 0.008$ \\
        \hline
    \end{tabular}
    \label{tab:c_dists}
\end{table*}

\begin{figure}
    \centering
    \includegraphics[width=\columnwidth]{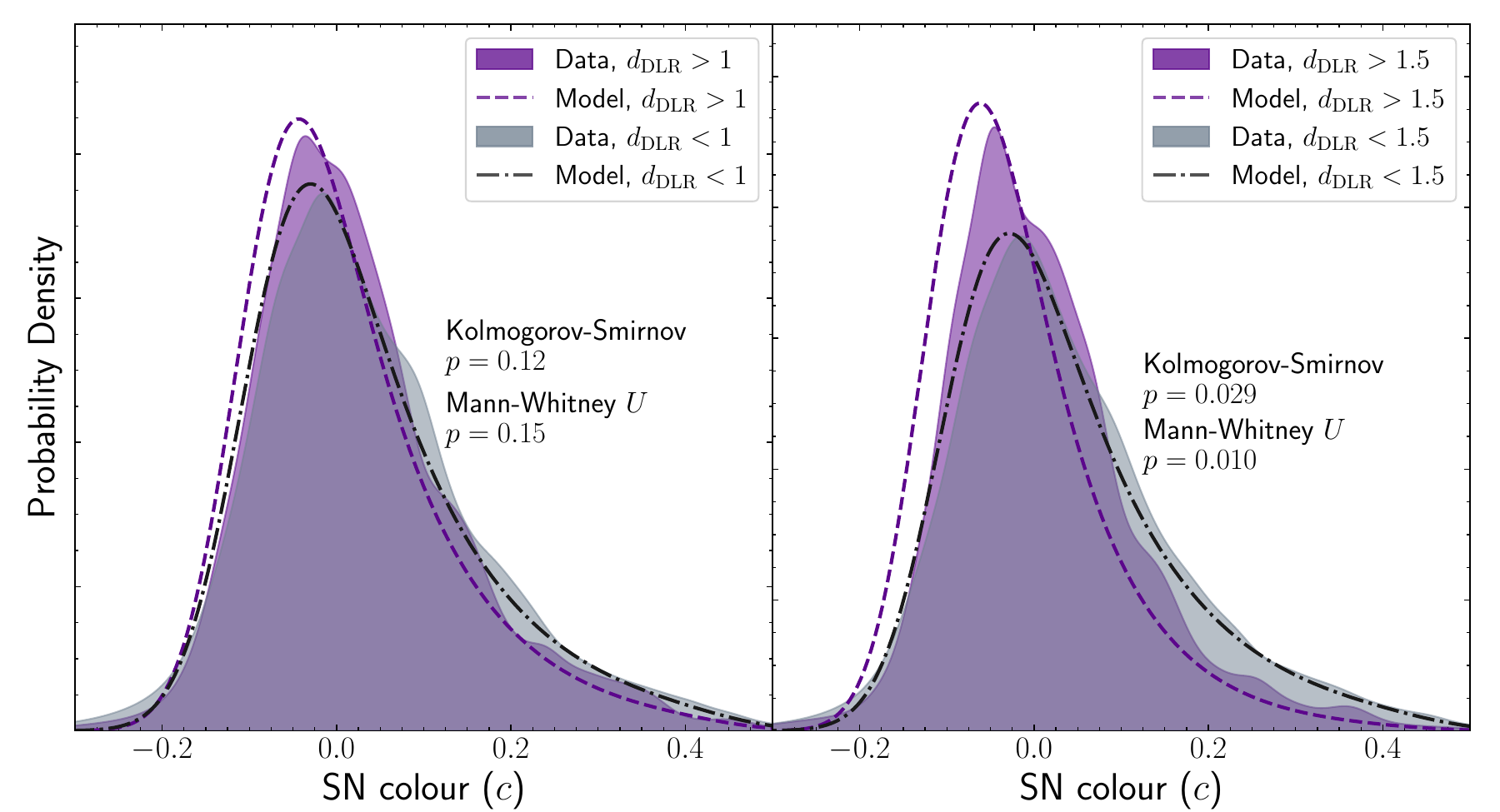}
    \caption{Left: SALT3 colour ($c$) distributions for SNe in inner ($\ddlr\le1$, grey) and outer ($\ddlr>1$, purple) regions, for a \lq colour test\rq\ sample with $z<0.6$ and SN $c<0.5$ . The overplotted models are the best-fitting intrinsic distributions, broadened by a uniform smear equal to the average uncertainty in the data. We find there is no significant difference between the samples.
    Right: As left, but when splitting at $\ddlr=1.5$. In this case there is a strong difference between the two colour distributions.}
    \label{fig:colour_hist}
\end{figure}

We perform two-sample statistical tests to determine whether the colour distributions are different. The SALT3 $c$ distributions for SNe in inner ($\ddlr\le1$) and outer ($\ddlr>1$) regions are consistent with being sampled from the same parent distribution: the two-sided Kolmogorov--Smirnov (KS) test statistic is 0.05 with a $p=0.22$, while a Mann--Whitney U test results in $p=0.10$. However, when splitting instead at $\ddlr=1.5$, we find {mild evidence that} the samples come from different populations (KS $p=0.03$, Mann--Whitey $U$ $p=0.008$).

To test whether this effect is caused by selection effects, we define a \lq colour test\rq\ sample. We limit the redshift to $z<0.6$ to retain higher quality light-curves, while relaxing the SN colour selection of the cosmological sample and allowing the reddest SNe Ia up to $c<0.5$ to probe the full reddening tail. {This sample is shown in Fig.~\ref{fig:colour_hist}}. This gives 1268 SNe Ia. We repeat the tests: {the two-sided KS test statistic is 0.05 with a $p=0.29$, while a Mann--Whitney U test results in $p=0.12$. However, when splitting instead at $\ddlr=1.5$, we find with high significance that the samples come from different populations (KS $p=5.6\times10^{-4}$, Mann--Whitey $U$ $p=1.9\times10^{-3}$).}

{The result that the colour distributions appear different when split at $\ddlr=1.5$, but not 1, both in the nominal DES-SN5YR and the \lq colour test\rq\ selections, is interesting}. This result indicates that SNe Ia at $\ddlr>1.5$ are less affected by reddening, and is consistent with the findings of \citet{2024arXiv240602072G}, but does not explain why a split of $\ddlr=1$ leads to a more significant difference in mass steps.

To further investigate these results we fit our data with the commonly used functional form as in \citet{2007ApJ...659..122J}, \citet{2011ApJ...731..120M}, and \citet{2021ApJ...909...26B}, where the intrinsic colour distribution is modelled by a Gaussian (mean $\mu_c$, width $\sigma_c$), with reddening introducing an exponential tail of scale $\tau_E$. This exponential follows the expected distribution of reddening within the disk of late-type galaxies \citep{1998ApJ...502..177H,2004NewAR..48..567C,2005MNRAS.362..671R}. We follow \citet{2024arXiv240602072G} in fitting the model directly to the data (without binning and fitting to a histogram), by evaluating the likelihood of each colour observation given the model and the uncertainty of the colour measurements, and minimizing the sum of the negative log-likelihood (equation~1 in \citealt{2024arXiv240602072G}).

The results are presented in Table \ref{tab:c_dists}, and the example of the model fitted to the \lq colour test\rq\ sample is shown in Fig.~\ref{fig:colour_hist}. {We note that on visual inspection the modes of the model differ from the data, and differ between samples. The plotted model requires the addition of an uncertainty ``smear" which we approximate by taking the median uncertainty of the data, but inevitably this leads to a slightly inaccurate representation of how the model fits the data.}
As above, when splitting the DES-SN5YR sample at $\ddlr=1$ we find $\mu_c$, $\sigma_c$ and $\tau_E$ are consistent between SNe Ia in the inner and outer regions, {which is different to what was found in} \citet{2024arXiv240602072G} where $\tau_E$ is reduced by $\sim 30$ per cent in the outer regions. {When limiting to the \lq colour test\rq\ sample, we again find that the model parameters are consistent between inner and outer regions when split at $\ddlr=1$, although for all \ddlr\ we find a significant shift to a redder $\mu_c$ and larger $\tau_E$ when compared to the full sample. However,} if we split at $\ddlr=1.5$ we find $\tau_E$ is significantly larger in outer regions, with the difference between inner and outer regions enhanced in the \lq colour test\rq\ sample. No matter our selection, the DES data are not fit by as large a $\tau_E$ as the low-$z$ ZTF \citet{2024arXiv240602072G} sample, due to selection effects, choice of light curve fitter, and possible redshift evolution. 

\subsubsection{Hubble residuals as a function of SN Ia colour}

\begin{figure*}
    \centering
    \includegraphics[width=\columnwidth]{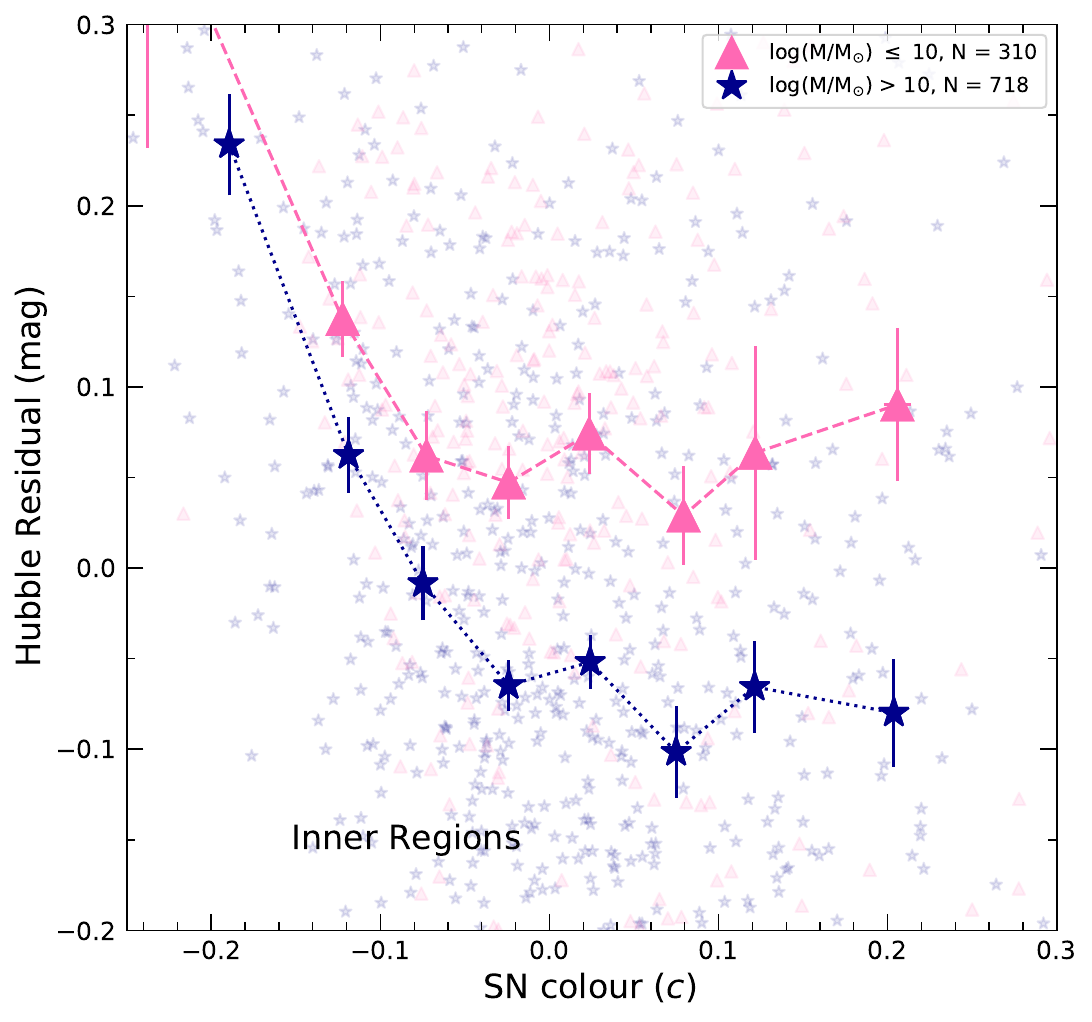}
	\includegraphics[width=\columnwidth]{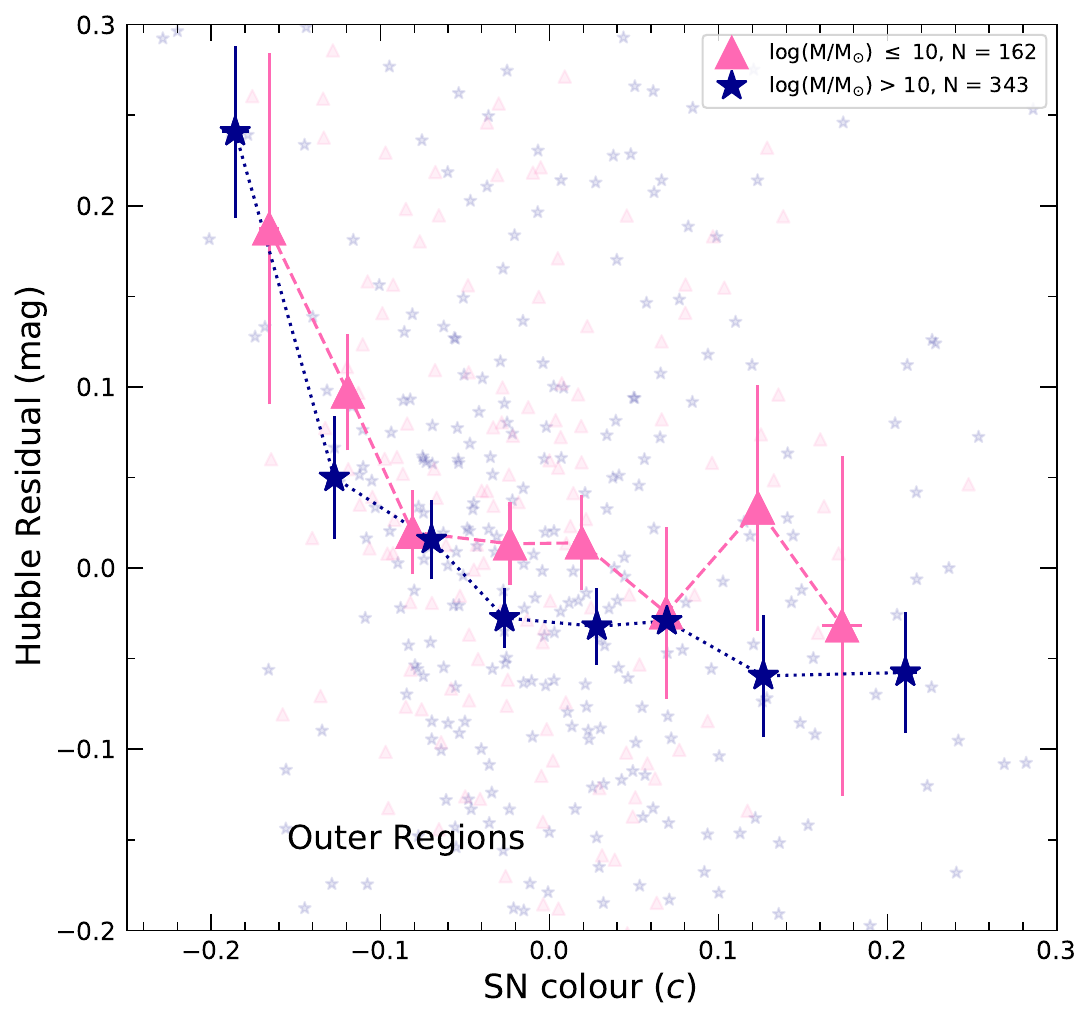}
    \caption{Hubble residuals for our inner (left) and outer (right) regions as a function of SALT3 colour $c$, plotted separately for events in low and high-mass host galaxies. The difference between the high- and low-mass points is effectively the mass step at any SN Ia colour.}
    \label{fig:crocandplot}
\end{figure*}

%Given the consistency of the SN Ia colour distributions between SNe Ia in inner and outer regions (Section~\ref{subsubsec:c_dists}),
We next consider the slope of the dust extinction laws as a possible reason for the difference in step size. The Hubble residuals are plotted against SN Ia colour for inner and outer regions in the left and right hand panels of Fig.~\ref{fig:crocandplot} respectively. Note that the apparent increase in blue SNe to a positive Hubble residual is a well-known selection bias \citep[see discussion in][]{2016ApJ...822L..35S} resulting from our use of a BBC 1D bias correction.

In inner regions, the mass step for all colours redder than $c=-0.1$ is clear, and increases with increasing colour: this is consistent with what was seen in the full DES-SN5YR sample \citep{2023MNRAS.519.3046K,2024arXiv240102945V}. The step in blue SNe Ia is not explained well by differing $R_V$ distributions and has been attributed to SNe Ia from differing ages of stellar population \citep{2023MNRAS.519.3046K,2022MNRAS.515.4587W,2024arXiv240605051P}. The step in blue SNe Ia in inner regions, but not in outer regions, may be explained by a more diverse stellar (and SN progenitor) population in the inner regions of galaxies, while the outer regions host a homogeneous population of SNe Ia regardless of the host mass. We can invoke a similar description for dust: inner regions evolve such that their average dust laws are different for host galaxies of different mass, while outer regions retain the same dust properties throughout the evolution of the galaxy as a whole. 

\begin{figure*}
    \centering
    \includegraphics[width=\textwidth]{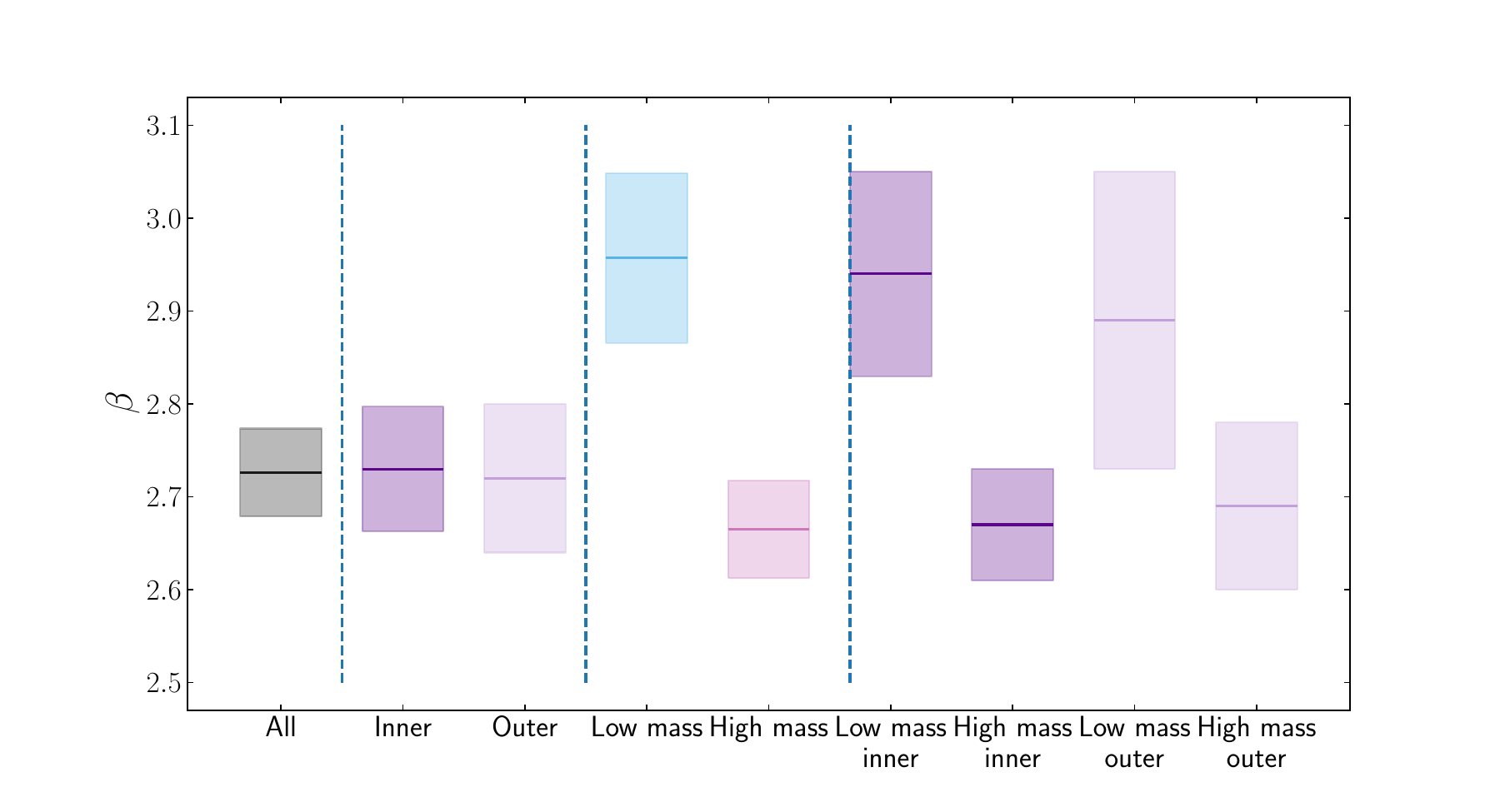}
    \caption{The strength of the colour--luminosity relationship parametrized by $\beta$, for different host galaxy selections. Our best-fitting estimates are represented by horizontal lines, with the $1\,\sigma$ uncertainty shown as a shaded region.}
    \label{fig:betas}
\end{figure*}

We additionally calculate the slope of the colour-luminosity relation ($\beta$) for each of our subsamples (split at $\ddlr=1$), shown in Fig.~\ref{fig:betas} and Table. \ref{tab:betas}. There is no overall difference between $\beta$ in inner and outer regions, but a significant difference ($\Delta \beta = 0.3$; $2.8\,\sigma$) between low and high-mass galaxies in general \citep{2010MNRAS.406..782S,2021ApJ...909...26B,2022ApJ...938...62C}. This difference is weakened ($\Delta \beta = 0.27$; $2.2\,\sigma$) in inner regions, and the significance is reduced further in outer regions ($\Delta \beta = 0.2$; $1.1\,\sigma$) as the sample size is much smaller. We find that SNe within the inner regions of their host are best fit with different $\beta$ values between high and low-mass hosts. Within the outer regions however, there is no significant difference between high and low-mass $\beta$.

\begin{figure*}
    \centering
    \includegraphics[width=\columnwidth]{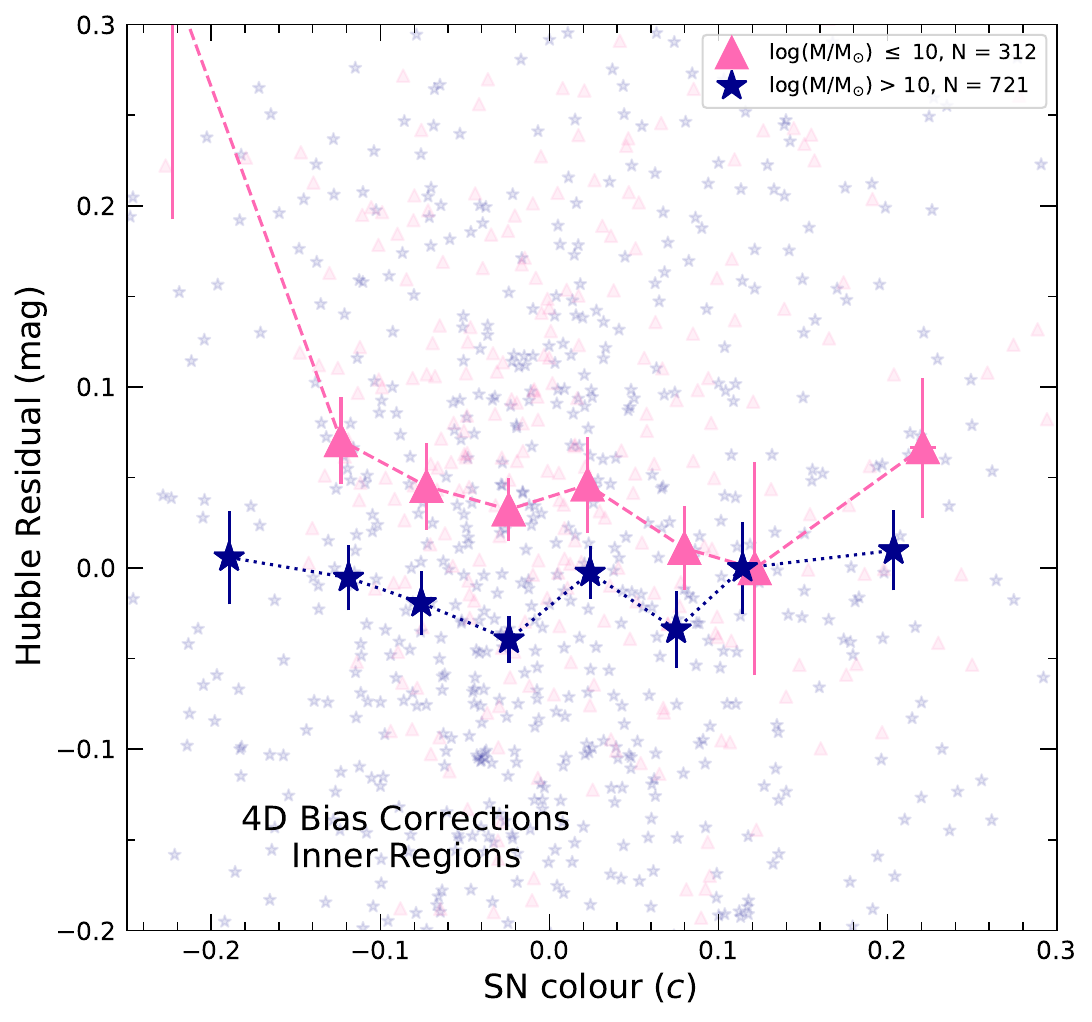}
	\includegraphics[width=\columnwidth]{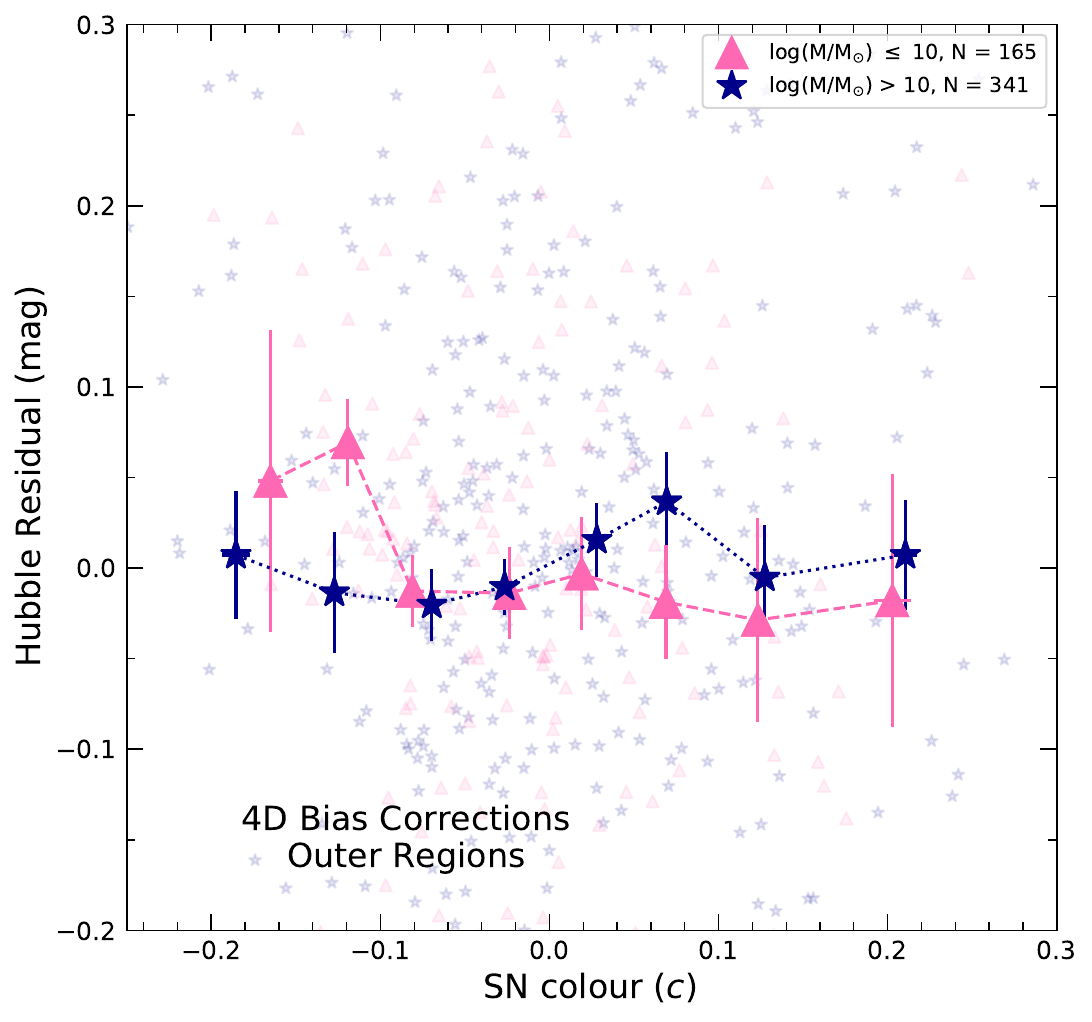}
    \caption{Hubble residuals for our inner (left) and outer (right) regions as a function of SALT3 colour $c$, plotted separately for events in low and high-mass host galaxies, after having made a BBC 4D correction for the colour-dependent selection and dust bias.}
    \label{fig:crocandplot_4D}
\end{figure*}

In Fig.~\ref{fig:crocandplot_4D} we show the standardized Hubble residuals after performing a BBC 4D bias correction which assumes a different $R_V$ distribution for low and high-mass galaxies. In inner regions, the characteristic curve of Hubble residual from blue to red SNe is removed, as is the differential difference between low and high-mass galaxies. However, a difference of $\sim 0.04$\,mag between low and high-mass hosts remains at all colours, as discussed in \citet{2024arXiv240102945V}. In outer regions, the curve is also removed but no significant residual difference exists at any SN colour.

\begin{table}
\caption{Strength of the SN Ia colour--luminosity relation $\beta$ (equation~\ref{eqn:mu}) for SNe Ia in different host galaxy environments.}
\centering
\begin{tabular}{lc}
\hline
Sample & $\beta$\\
\hline
Full Sample& $2.72 \pm 0.05$\\
$d_{\mathrm{DLR}} \leq 1$ & $2.72 \pm 0.08$ \\
$d_{\mathrm{DLR}}> 1$ & $2.73 \pm 0.07$ \\
$\log(M_*/\msolar)\leq 10$ & $2.96 \pm 0.09$ \\
$\log(M_*/\msolar)> 10$ & $2.66 \pm 0.05$ \\
\hline
\multicolumn{2}{c}{$d_{\mathrm{DLR}} \leq 1$}  \\
$\log(M_*/\msolar)\leq 10$ & $2.94 \pm 0.11$ \\
$\log(M_*/\msolar)> 10$ & $2.67 \pm 0.06$ \\
\hline
\multicolumn{2}{c}{$d_{\mathrm{DLR}} > 1$}  \\
$\log(M_*/\msolar)\leq 10$ & $2.89 \pm 0.16$ \\
$\log(M_*/\msolar)> 10$ & $2.69 \pm 0.09$ \\
\hline
\end{tabular}
\label{tab:betas}

\end{table}

\subsection{Populations of SN Ia progenitors}

A difference in the effects of dust extinction between inner and outer regions does not seem to be the cause of the difference in host galaxy steps between those regions. A different explanation could be that the SN explosions themselves occur in multiple types of progenitors, with a different average standardized peak brightness. These varied populations of progenitor systems could then be present in differing proportions between the inner regions of low-mass/blue/young and high-mass/red/old galaxies, but the outer regions of all galaxies only host a single population.

These progenitors types are often assumed to relate to the \lq delay time\rq\ from an initial burst of star formation to the SN Ia explosion. For the SN Ia population as a whole, the distribution of these delay times -- the delay time distribution (DTD) -- is well described by a declining power law \citep{2004MNRAS.347..951M,2008PASJ...60.1327T,2010ApJ...722.1879M,2014ARA&A..52..107M,2015MNRAS.450..905G,2021MNRAS.501.3122C,2021MNRAS.506.3330W}. When such a DTD is convolved with the stellar age distribution of stars across different galaxy types, a bimodal distribution for the age of SN progenitors at the time of explosion is observed \citep{2005A&A...433..807M,2005ApJ...629L..85S,2006MNRAS.370..773M,2006ApJ...648..868S,2012ApJ...755...61S,2014MNRAS.445.1898C,2021MNRAS.506.3330W}: so-called \lq prompt\rq\ and \lq delayed \rq\ components.

\citet{2013A&A...560A..66R} and \citet{2014MNRAS.445.1898C} proposed that if a difference in average peak brightness of the prompt and delayed populations is the cause of the step, then the step size should change as a function of redshift. Evidence for such an evolution is limited, although we note the step observed in the high-$z$ DES-SN5YR sample (before dust-like bias corrections) is somewhat smaller than that seen in the low-$z$ ZTF sample \citep{2024arXiv240602072G}. There is, however, strong evidence that these populations do have different light curve characteristics. Their stretch distributions are different, with faster declining light curves (lower $x_1$) occurring more frequently in environments with older, more passive stellar populations \citep[e.g.,][]{1995AJ....109....1H, 2006ApJ...648..868S}.

Recent analyses observe two distinct \lq modes\rq\ of SN stretch: a low-stretch mode which is observed only in passive environments, and a high-stretch mode observed in both star-forming and passive environments \citep[e.g.,][]{2020A&A...644A.176R}. \citet{2021A&A...649A..74N} showed that the relative abundance of these modes in an observed population of SNe Ia evolves as a function of redshift, which can be explained if the modes relate to the prompt and delayed modes of the SN Ia DTD. What is not clear is whether the two modes have different standardized peak brightnesses, or require different standardization parameters $\alpha$ and $\beta$. 

\begin{figure*}
    \centering
    \includegraphics[width=0.49\textwidth]{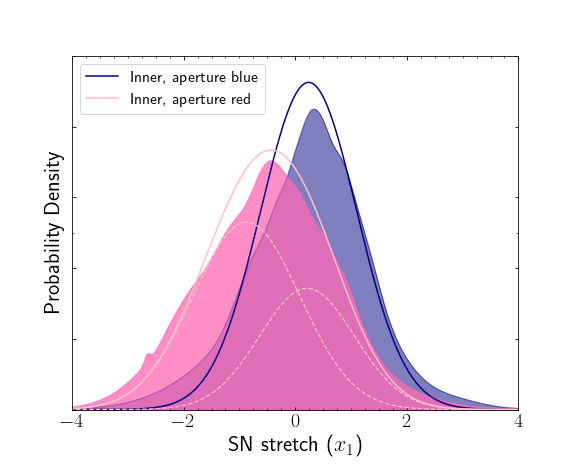}
	\includegraphics[width=0.49\textwidth]{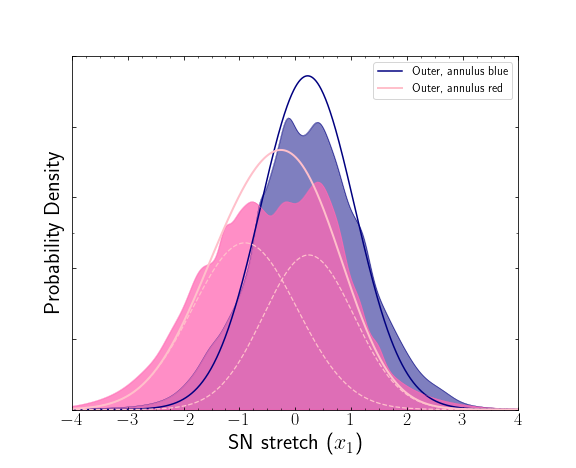}
    \caption{SN stretch distributions for the inner ($\ddlr\le1$) and outer ($\ddlr>1$) regions, split according to the colour of the aperture/annulus of the location of the SN. The models plotted are results of fit directly to the data, and are drawn by adding a uniform smear equal to the mean $x_1$ uncertainty to the intrinsic width.}
    \label{fig:x1_dists}
\end{figure*}

We test whether there is a difference in the SN Ia populations between the inner and outer regions by modelling the $x_1$ distributions according to the \citet{2021A&A...649A..74N} model. By using the annulus colour as a proxy for the age of the stellar population, we fix their $y^i$ parameter, which is the probability that the progenitor is young, according to
\begin{equation}
    y^i = \left\{
    \begin{array}{@{}rc@{}}
    1, & (U-R)_{\mathrm{annulus}} <1\\
    0 , & (U-R)_{\mathrm{annulus}}\geq 1 \\
    \end{array}\right. \,.
\end{equation}
The model for the $x_1$ distribution is then 
\begin{equation}
    p(x_1) \sim \left\{
    \begin{array}{@{}ll@{}}
    \mathcal{N}(\mu_{1,x_1},\sigma_{1,x_1}), & (U-R)_{\mathrm{annulus}} <1\\
   a\mathcal{N}(\mu_{1,x_1},\sigma_{1,x_1}) + & \\
   (1-a)\mathcal{N}(\mu_{2,x_1},\sigma_{2,x_1}), & (U-R)_{\mathrm{annulus}} >1\\
    \end{array}\right. \,. 
\end{equation}
The parameter $a$ determines the fraction of high-stretch mode SNe occurring in the delayed stellar population. 

We fit the full dataset with this model by minimizing the negative log-likelihood, and show the data and best-fitting model in Fig \ref{fig:x1_dists}. {We verify that the two Gaussian model is preferred over a single Gaussian via the Akaike information criterion: taking the significance of the preference as $\exp{\Delta\mathrm{AIC}/2}$, we find that the double Gaussian is preferred to $6.8 \sigma$. We do not use a likelihood ratio test since the models are not strictly nested.} We find $\mu_{1,x_1}=0.22\pm0.03$, $\sigma_{1,x_1}=0.67\pm0.03$, $\mu_{2,x_1}=-0.86\pm0.17$, $\sigma_{2,x_1}=0.80\pm0.07$, and $a=0.37\pm0.1$ which are broadly consistent with the results of \citet{2021A&A...649A..74N} and \citet{2024arXiv240520965G} except for $a$ which is significantly smaller than found in those works but consistent with the value for DES-SN5YR modelled in \citet{2022MNRAS.515.4587W}. This small $a$ could be caused by our fixing of $y^i$ based on the galaxy colours: it is known that no galaxy observable perfectly traces the underlying properties, and even locally measured colours experience contamination whereby young SNe occur in populations measured to be old \citep{2022A&A...657A..22B}. If instead we fix the probability that any given SN is young to the average for our mean redshift according to the \citet{2020A&A...644A.176R} drift model, $\delta(z=0.53)=0.67$, we find  $\mu_{1,x_1}=0.06$, $\sigma_{1,x_1}=0.70$, $\mu_{2,x_1}=-1.65$, $\sigma_{2,x_1}=0.45$, and $a=0.58$. Note that these mean values are slightly more negative than the low-$z$ values from \citet{2021A&A...649A..74N} and \citet{2024arXiv240520965G} -- this could be genuine redshift evolution or an effect of the use of SALT2 rather than SALT3 light curve fits in those works.

Next, we fit the inner and outer SNe separately, fixing the model parameters to those found using the aperture/annulus colour as a proxy for stellar population age, except for $a$ which we leave free. We find $a=0.32\pm0.05$ in inner regions and $a=0.42\pm0.04$ in outer regions, which means there {may be} a larger fraction of high-stretch mode SNe in red outer annuli compared to red central regions. 

The results of this test indicate that inner regions are more strongly split between the high-stretch SN mode in young and blue stellar populations and the low-stretch SN mode in old and red stellar populations. The outer regions have a more mixed population of low and high-stretch modes in their old and red stellar populations. If the Hubble residual step is caused by differences between these two modes, then this difference could be a tangible explanation. In particular, \citet{2022MNRAS.515.4587W} noted that the inferred strength of the width-luminosity relation $\alpha$ is degenerate with a Hubble residual step that is related to the stellar age of the SN progenitor. If $\alpha$ is being mis-measured because it actually has a different strength for the two modes, this could introduce an artificial Hubble residual dependence that is related to the stellar population age.

However, \citet{2024arXiv240520965G} modelled their SN distances, standardized for colour but not stretch, with a two-component \lq broken\rq $\alpha$ and found that while the standardization is improved, the host mass or host colour Hubble residual step \emph{increases}. Furthermore, the difference between the Hubble residuals in inner and outer regions appears more driven by the low-mass and blue environments whose Hubble residuals become more negative in outer regions, while the high-mass and red environments are relatively unchanged. That is, the difference in Hubble residuals is insensitive to the relative fraction of low/high-stretch SNe in red/high-mass hosts.

\subsection{Implications for cosmological measurements}
Distance measurements using SNe Ia are at the forefront of cosmology due to an increase in sample sizes combined with an improved handling of systematics. In DES-SN5YR, the systematic uncertainty on the dark energy equation of state parameter $w$ is smaller than the statistical uncertainty \citep{2024arXiv240102945V}. Nevertheless, the largest component of this systematic uncertainty is the colour-and-host dependent scatter model and associated bias corrections. Given that the values used for the BBC-4D bias correction are conditioned on the full dataset, it could be that the corrections are underestimated in inner regions, and overestimated in outer regions. A full analysis of the impact of this effect is beyond the scope of this work. Here we note that for the effect to propagate through to a bias on $w$, it would require the distribution of \ddlr \hspace{0.03cm} to evolve with redshift. Instead the \ddlr \hspace{0.03cm} distribution remains constant as a function of redshift in DES-SN5YR.
\section{Summary}

Using the DES five-year sample of photometrically-classified SNe Ia, we have analysed the relationships between projected and normalized host galaxy separation and the SN Ia light-curve properties and inferred cosmological distance measurements. Using the directional light radius (DLR) method as our normalisation of the galactocentric distances (\ddlr), our main findings are:

\begin{itemize}
    \item We confirm previous findings of \citet{2023MNRAS.526.5292T} that the innermost regions of galaxies (\ddlr $\leq 1$) host faster-declining SNe than the outer regions. We also confirm earlier results concerning SN Ia colour, and show that SNe Ia in the outer regions of galaxies ($\ddlr>1.5$) are bluer ($\bar{c}=-0.023\pm0.006$) than those at $\ddlr\leq1.5$ ($\bar{c}=0.010\pm0.003$), a difference of $0.033\pm0.007$ ($\simeq4.7\sigma$).
    \item We show, for the first time, that the difference in SN Ia post-standardization brightnesses between high and low-mass hosts reduces from $0.078\pm0.011$\,mag in the full sample to $0.036 \pm 0.018$\,mag for SNe Ia located in the outer regions of their host galaxies ($\ddlr>1$), while increasing to $0.100 \pm 0.014$\,mag for SNe in the inner regions ($\ddlr\le1$). The difference in the size of the mass step between inner and outer regions is $0.064\pm0.023$\,mag. The effect remains when splitting SNe Ia by their global galaxy $U-R$ colour, or by the $U-R$ colour in the inner aperture/outer annulus in which the SN occurred.
    \item We find that the decrease in magnitude of this mass step as a function of \ddlr\ is well fit by either a step function, split at \ddlr\ of 1, or a linear function with \ddlr.
    \item We show that using different $R_V$ values for dust along the line of sight to SNe that varies between low and high-mass host galaxies, can reduce but not remove the step for SNe Ia in the inner regions. 
    \item There is no evidence that the outer regions of galaxies have dust laws affecting SNe Ia that change as a function of {the galaxy's} stellar mass. Similarly, there is no evidence for an intrinsic luminosity difference between SNe Ia in the outer regions of low and high-mass galaxies.
    \item We find a slight difference between the strength of the high-stretch mode of SN Ia $x_1$ in red inner and red outer regions, but consider this effect unlikely to be the cause of the Hubble residual effect.
\end{itemize}

Calculating \ddlr\ is an algorithmically  straight forward task, and selecting SNe Ia based on \ddlr\ reduces the need to account for host galaxy properties to standardize SN Ia brightnesses across different galaxies, without much computational effort. While we have not identified the underlying astrophysics that such projected galactocentric distances are tracing in the DES five-year SN Ia sample, we have shown that the standardized distance measurements from SNe Ia in the outer regions of galaxies have {a reduced} dependence on their global host galaxy properties. 

Restricting a cosmological analysis to SNe Ia in the outer regions of their host galaxies reduces the sample size to around a third. This will increase the statistical uncertainties. However, the reduction in the astrophysical systematic uncertainties (and complications) gained from using such a sample, coupled with the very large sample sizes expected in future experiments such as the Rubin Observatory's Legacy Survey of Space and Time (LSST) \citep{Ivezic2008}, means that such a selection is likely to be {a useful validation of the results found using a full sample.}

\section*{Acknowledgements}
We thank the referee for a thoughtful review.
All authors have contributed to the drafting of this manuscript. MT devised the project and led the analysis. PW and MS provided scientific guidance and support with the production of the manuscript. MS contributed host galaxy fitting and wrote substantial sections of the manuscript, and PW ran 1D bias corrections and provided the colour and stretch analyses. DS and MV internally reviewed the work and provided extensive feedback. CF, CL, JL, LK, LG, RK, PS, and TD provided comments on the analysis and interpretation. All aforementioned authors as well as AM, BP, BS, DB, and MS contributed to the DES-SN5YR data and methods used in this paper. The remaining authors have made contributions to this paper that include, but are not limited to, the construction of DECam and other aspects of collecting the data; data processing and calibration; developing broadly used methods, codes, and simulations; running the pipelines and validation tests; and promoting the science analysis.

PW and MS acknowledge support from the Science and Technology Facilities Council (STFC) grants ST/R000506/1 and ST/Y001850/1. 

This work was completed in part with resources provided by the University of Chicago’s Research Computing Center.

Funding for the DES Projects has been provided by the U.S. Department of Energy, the U.S. National Science Foundation, the Ministry of Science and Education of Spain, 
the Science and Technology Facilities Council of the United Kingdom, the Higher Education Funding Council for England, the National Center for Supercomputing 
Applications at the University of Illinois at Urbana-Champaign, the Kavli Institute of Cosmological Physics at the University of Chicago, 
the Center for Cosmology and Astro-Particle Physics at the Ohio State University,
the Mitchell Institute for Fundamental Physics and Astronomy at Texas A\&M University, Financiadora de Estudos e Projetos, 
Funda{\c c}{\~a}o Carlos Chagas Filho de Amparo {\`a} Pesquisa do Estado do Rio de Janeiro, Conselho Nacional de Desenvolvimento Cient{\'i}fico e Tecnol{\'o}gico and 
the Minist{\'e}rio da Ci{\^e}ncia, Tecnologia e Inova{\c c}{\~a}o, the Deutsche Forschungsgemeinschaft and the Collaborating Institutions in the Dark Energy Survey. 

The Collaborating Institutions are Argonne National Laboratory, the University of California at Santa Cruz, the University of Cambridge, Centro de Investigaciones Energ{\'e}ticas, 
Medioambientales y Tecnol{\'o}gicas-Madrid, the University of Chicago, University College London, the DES-Brazil Consortium, the University of Edinburgh, 
the Eidgen{\"o}ssische Technische Hochschule (ETH) Z{\"u}rich, 
Fermi National Accelerator Laboratory, the University of Illinois at Urbana-Champaign, the Institut de Ci{\`e}ncies de l'Espai (IEEC/CSIC), 
the Institut de F{\'i}sica d'Altes Energies, Lawrence Berkeley National Laboratory, the Ludwig-Maximilians Universit{\"a}t M{\"u}nchen and the associated Excellence Cluster Universe, 
the University of Michigan, NSF NOIRLab, the University of Nottingham, The Ohio State University, the University of Pennsylvania, the University of Portsmouth, 
SLAC National Accelerator Laboratory, Stanford University, the University of Sussex, Texas A\&M University, and the OzDES Membership Consortium.

Based in part on observations at NSF Cerro Tololo Inter-American Observatory at NSF NOIRLab (NOIRLab Prop. ID 2012B-0001; PI: J. Frieman), which is managed by the Association of Universities for Research in Astronomy (AURA) under a cooperative agreement with the National Science Foundation.

The DES data management system is supported by the National Science Foundation under Grant Numbers AST-1138766 and AST-1536171.
The DES participants from Spanish institutions are partially supported by MICINN under grants PID2021-123012, PID2021-128989 PID2022-141079, SEV-2016-0588, CEX2020-001058-M and CEX2020-001007-S, some of which include ERDF funds from the European Union. IFAE is partially funded by the CERCA program of the Generalitat de Catalunya.

We  acknowledge support from the Brazilian Instituto Nacional de Ci\^encia
e Tecnologia (INCT) do e-Universo (CNPq grant 465376/2014-2).

This manuscript has been authored by Fermi Research Alliance, LLC under Contract No. DE-AC02-07CH11359 with the U.S. Department of Energy, Office of Science, Office of High Energy Physics.

\section*{Data Availability}

All data used in this article are publicly available with the DES-SN5YR data release \citep{2024arXiv240605046S} at \url{https://github.com/des-science/DES-SN5YR}.

\bibliographystyle{mnras}
\bibliography{paper_draft} 
\appendix
\section{BBC1D distances}
{In Table \ref{tab:ancilliary} we present the `BBC1D' corrected distance moduli for our 1531 DES5YR SNe. A full version of the table can be found online. Full data for each SN can be found in the DES5YR data release described in \citet{2024arXiv240605046S} and found at \url{https://des-sn-dr.readthedocs.io/en/latest/}.}
\begin{table}
    \centering
    \begin{tabular}{|c|c|c|c|}
        \hline
         CID$^{\mathrm{a}}$& $\mu^{\mathrm{b}}$ & $\Delta\mu^{\mathrm{c}}$  & $\mu_{\mathrm{bias}}^{\mathrm{d}}$\\
         \hline
         1702083 & 41.7034 &  0.1041 &     -0.0050 \\
         1435094 & 43.6092 & -0.5358 &     -0.0697 \\
         1309227 & 44.2862 &  0.3922 &     -0.0472 \\
         1324542 & 42.5540 &  0.1201 &     -0.0325 \\
         1320474 & 43.6429 &  0.2471 &     -0.1081 \\
    \end{tabular}
    \caption{a) DES unique SN identifier; b) Distance modulus from Eq. \ref{eqn:mu}; c) Residual of $\mu$ with respect to the fiducial $\Lambda$CDM cosmological model (Eq \ref{eq:hubbleresiduals}); d) Size of the 1D bias correction applied to achieve the $\mu$ estimate.}
    \label{tab:ancilliary}
\end{table}
\onecolumn
\parbox{\textwidth}{
%\scriptsize
$^{1}$ School of Physics and Astronomy, University of Southampton,  Southampton, SO17 1BJ, UK\\
$^{2}$ Department of Physics, Duke University Durham, NC 27708, USA\\
$^{3}$ Einstein Fellow\\
$^{4}$ Department of Physics, University of Oxford, Denys Wilkinson Building, Keble Road, Oxford OX1 3RH, United Kingdom\\
$^{5}$ Center for Astrophysics $\vert$ Harvard \& Smithsonian, 60 Garden Street, Cambridge, MA 02138, USA\\
$^{6}$ School of Mathematics and Physics, University of Queensland,  Brisbane, QLD 4072, Australia\\
$^{7}$ Institut d'Estudis Espacials de Catalunya (IEEC), 08034 Barcelona, Spain\\
$^{8}$ Institute of Space Sciences (ICE, CSIC),  Campus UAB, Carrer de Can Magrans, s/n,  08193 Barcelona, Spain\\
$^{9}$ Centre for Gravitational Astrophysics, College of Science, The Australian National University, ACT 2601, Australia\\
$^{10}$ The Research School of Astronomy and Astrophysics, Australian National University, ACT 2601, Australia\\
$^{11}$ Department of Physics and Astronomy, University of Pennsylvania, Philadelphia, PA 19104, USA\\
$^{12}$ Institute of Cosmology and Gravitation, University of Portsmouth, Portsmouth, PO1 3FX, UK\\
$^{13}$ Department of Astronomy and Astrophysics, University of Chicago, Chicago, IL 60637, USA\\
$^{14}$ Kavli Institute for Cosmological Physics, University of Chicago, Chicago, IL 60637, USA\\
$^{15}$ Centre for Astrophysics \& Supercomputing, Swinburne University of Technology, Victoria 3122, Australia\\
$^{16}$ Universit\'e Grenoble Alpes, CNRS, LPSC-IN2P3, 38000 Grenoble, France\\
$^{17}$ Department of Physics \& Astronomy, University College London, Gower Street, London, WC1E 6BT, UK\\
$^{18}$ Physics Department, Lancaster University, Lancaster, LA1 4YB, UK\\
$^{19}$ Laborat\'orio Interinstitucional de e-Astronomia - LIneA, Av. Pastor Martin Luther King Jr, 126 Del Castilho, Nova Am\'erica Offices, Torre 3000/sala 817 CEP: 20765-000, Brazil\\
$^{20}$ Fermi National Accelerator Laboratory, P. O. Box 500, Batavia, IL 60510, USA\\
$^{21}$ Department of Physics, University of Michigan, Ann Arbor, MI 48109, USA\\
$^{22}$ Kavli Institute for Particle Astrophysics \& Cosmology, P. O. Box 2450, Stanford University, Stanford, CA 94305, USA\\
$^{23}$ SLAC National Accelerator Laboratory, Menlo Park, CA 94025, USA\\
$^{24}$ Instituto de Astrofisica de Canarias, E-38205 La Laguna, Tenerife, Spain\\
$^{25}$ Universidad de La Laguna, Dpto. Astrofísica, E-38206 La Laguna, Tenerife, Spain\\
$^{26}$ Institut de F\'{\i}sica d'Altes Energies (IFAE), The Barcelona Institute of Science and Technology, Campus UAB, 08193 Bellaterra (Barcelona) Spain\\
$^{27}$ Hamburger Sternwarte, Universit\"{a}t Hamburg, Gojenbergsweg 112, 21029 Hamburg, Germany\\
$^{28}$ Department of Physics, IIT Hyderabad, Kandi, Telangana 502285, India\\
$^{29}$ California Institute of Technology, 1200 East California Blvd, MC 249-17, Pasadena, CA 91125, USA\\
$^{30}$ Institute of Theoretical Astrophysics, University of Oslo. P.O. Box 1029 Blindern, NO-0315 Oslo, Norway\\
$^{31}$ Instituto de Fisica Teorica UAM/CSIC, Universidad Autonoma de Madrid, 28049 Madrid, Spain\\
$^{32}$ Center for Astrophysical Surveys, National Center for Supercomputing Applications, 1205 West Clark St., Urbana, IL 61801, USA\\
$^{33}$ Department of Astronomy, University of Illinois at Urbana-Champaign, 1002 W. Green Street, Urbana, IL 61801, USA\\
$^{34}$ Santa Cruz Institute for Particle Physics, Santa Cruz, CA 95064, USA\\
$^{35}$ Center for Cosmology and Astro-Particle Physics, The Ohio State University, Columbus, OH 43210, USA\\
$^{36}$ Department of Physics, The Ohio State University, Columbus, OH 43210, USA\\
$^{37}$ Australian Astronomical Optics, Macquarie University, North Ryde, NSW 2113, Australia\\
$^{38}$ Lowell Observatory, 1400 Mars Hill Rd, Flagstaff, AZ 86001, USA\\
$^{39}$ Jet Propulsion Laboratory, California Institute of Technology, 4800 Oak Grove Dr., Pasadena, CA 91109, USA\\
$^{40}$ George P. and Cynthia Woods Mitchell Institute for Fundamental Physics and Astronomy, and Department of Physics and Astronomy, Texas A\&M University, College Station, TX 77843,  USA\\
$^{41}$ LPSC Grenoble - 53, Avenue des Martyrs 38026 Grenoble, France\\
$^{42}$ Instituci\'o Catalana de Recerca i Estudis Avan\c{c}ats, E-08010 Barcelona, Spain\\
$^{43}$ Department of Physics, Carnegie Mellon University, Pittsburgh, Pennsylvania 15312, USA\\
$^{44}$ Observat\'orio Nacional, Rua Gal. Jos\'e Cristino 77, Rio de Janeiro, RJ - 20921-400, Brazil\\
$^{45}$ Department of Physics and Astronomy, Pevensey Building, University of Sussex, Brighton, BN1 9QH, UK\\
$^{46}$ Department of Physics, Northeastern University, Boston, MA 02115, USA\\
$^{47}$ Centro de Investigaciones Energ\'eticas, Medioambientales y Tecnol\'ogicas (CIEMAT), Madrid, Spain\\
$^{48}$ Physik-Institut, University of Zürich, Winterthurerstrasse 190, CH-8057 Zürich, Switzerland\\
$^{49}$ Computer Science and Mathematics Division, Oak Ridge National Laboratory, Oak Ridge, TN 37831\\
$^{50}$ High Energy Physics Division, Argonne National Laboratory, 9700 S. Cass Avenue, Argonne, IL 60439, USA\\
$^{51}$ Cerro Tololo Inter-American Observatory, NSF's National Optical-Infrared Astronomy Research Laboratory, Casilla 603, La Serena, Chile\\
$^{52}$ Department of Astronomy, University of California, Berkeley,  501 Campbell Hall, Berkeley, CA 94720, USA\\
$^{53}$ Lawrence Berkeley National Laboratory, 1 Cyclotron Road, Berkeley, CA 94720, USA\\
}

\bsp	% typesetting comment
\label{lastpage}
\end{document}